\documentclass[11pt]{article}
\usepackage{amssymb,amsmath,amsfonts}
\usepackage{graphicx}
\usepackage{graphics}
\usepackage{eepic,epsfig}

\@addtoreset{equation}{section}

\makeatother

\begin{document}

\title{Finite temperature fermionic condensate in a conical space with a
circular boundary and magnetic flux}
\author{Aram A. Saharian$^{1}$\thanks{
E-mail: saharian@ysu.am}, Eugenio R. Bezerra de Mello$^{2}$\thanks{
E-mail: emello@fisica.ufpb.br}, Astghik A. Saharyan$^{3}$\thanks{%
astghik.saharyan@gmail.com} \vspace{0.3cm} \\
\textit{$^1$Department of Physics, Yerevan State University,}\\
\textit{1 Alex Manoogian Street, 0025 Yerevan, Armenia} \vspace{0.3cm}\\
\textit{$^{2}$Departamento de F\'{\i}sica, Universidade Federal da Para\'{\i}
ba}\\
\textit{58.059-970, Caixa Postal 5.008, Jo\~{a}o Pessoa, PB, Brazil}\vspace{%
	0.3cm}\\
\textit{$^3$Laboratoire Interdisciplinaire Carnot de Bourgogne, CNRS UMR
6303,}\\
\textit{Universit\'{e} Bourgogne Franche-Comt\'{e}, BP 47870, F-21078 Dijon,
France} \vspace{0.3cm}}
\maketitle

\begin{abstract}
We investigate the edge effects on the finite temperature fermionic
condensate (FC) for a massive fermionic field in a (2+1)-dimensional conical
spacetime with a magnetic flux located at the cone apex. The field obeys the
bag boundary condition on a circle concentric with the apex. The analysis is
presented for both the fields realizing two irreducible representations of
the Clifford algebra and for general case of the chemical potential. In both
the regions outside and inside the circular boundary, the FC is decomposed
into the boundary-free and boundary-induced contributions. They are 
even functions under the simultaneous change of the signs for the magnetic
flux and the chemical potential. The dependence of the FC on the magnetic
flux becomes weaker with decreasing planar angle deficit. For points near
the boundary, the effects of finite temperature, of planar angle deficit and
of magnetic flux are weak. For a fixed distance from the boundary and at
high temperatures the FC is dominated by the Minkowskian part. The FC in
parity and time-reversal symmetric (2+1)-dimensional fermionic models is
discussed and applications are given to graphitic cones.
\end{abstract}

\section{Introduction}

The investigation of field theoretical models in spatial dimensions $D$
other than 3 is motivated by several reasons. Many of high-energy theories
unifying physical interactions are formulated in higher-dimensional spaces, $%
D>3$, with compactified extra dimensions. Examples of this kind of models
include Kaluza-Klein theories, supergravity and various types of string/M
theories. In recent years the models with $D<3$ has attracted a great deal
of attention. Because of lower dimension they are easier to handle and are
treated as simplified models in particle physics. The models in dimensions $%
D<3$ appear as high temperature limits of $D=3$ field theories and also as
effective theories describing the long-wavelength dynamics of excitations in
condensed matter systems \cite{Naga99}-\cite{Mari17}. The examples for this
kind of systems include graphene made structures (such as a graphene sheet,
carbon nanotubes and nanoloops, nanoribbons), topological insulators, Weyl
semimetals and high-temperature superconductors. In the continuous limit,
the long-wavelength properties of these systems are well described by the
Dirac equation for fermionic fields living in (2+1)-dimensional spacetime
with the Fermi velocity instead of the velocity of light \cite{Gusy07}-\cite%
{Xiao11}. This offers the remarkable possibility to probe field theoretical
effects in condensed matter systems. Interesting features in
(2+1)-dimensional models include fractionalization of quantum numbers, the
possibility of the excitations with fractional statistics, flavour symmetry
breaking and parity violation. In the corresponding gauge theories, the
presence of topologically non-trivial gauge invariant terms in the action
provides an interesting possibility to give masses for gauge bosons \cite%
{Dese82,Niem83}. The infrared cutoff induced by the topological mass term
provides a way to solve the infrared problem without changing the
ultraviolet behavior.

In the present paper we investigate the combined effects of nontrivial
topology, induced by a conical defect, and of a circular boundary on the
finite temperature fermionic condensate (FC) for a massive fermionic field
in (2+1)-dimensional spacetime. In order to have an exactly solvable problem
in the region inside the boundary, a simplified model for the defect will be
used with a point-like core. The FC is among the most important local
characteristics of a given state for a fermionic field. It carries also an
important information about the global properties of the background
spacetime. The FC plays a central role in the models of dynamical breaking
of chiral symmetry (for chiral symmetry breaking in models with nontrivial
topology and in curved specitimes see \cite{Buch89}). The vacuum FC, the
vacuum expectation values of the charge and current densities and of the
energy-momentum tensor for a fermionic field in the geometry under
consideration have been investigated in \cite{Bell11}-\cite{Beze12}. The
finite temperature effects for the FC and for the charge and current
densities in a boundary-free (2+1)-dimensional conical space were discussed
in \cite{Bell16T} (for the corresponding effects in (3+1)-dimensional
spacetime with a cosmic string see \cite{Moha15}).

The (3+1)-dimensional analog of the setup we are going to consider here is
the geometry of an infinite straight cosmic string with coaxial cylindrical
boundary. The combined effects of the topology and boundary on the
properties of the quantum vacuum in that geometry have been considered for
electromagnetic \cite{Brev95,Beze07,Nest11}, scalar \cite{Nest11,Beze06sc}
and fermionic \cite{Beze08} fields. The Casimir forces for massless scalar
fields with Dirichlet and Neumann boundary conditions in the geometry of a
conical piston are investigated in \cite{Fucc11}. The scalar and
electromagnetic Casimir densities in the presence of boundaries
perpendicular to the string axis are discussed in \cite{Beze11}-\cite{Beze18}%
. Another type of boundary conditions on quantum fields arise for a cosmic
strings compactified along its axis. The influence of the compactification
on the properties of the quantum vacuum were investigated in \cite%
{Beze12Comp}.

The organization of the paper is as follows. In the next section we describe
the bulk and boundary geometries, the field and present complete sets of
fermionic modes outside and inside a circular boundary. By using those
modes, in section \ref{sec:Ext}, the FC in the exterior region is evaluated.
It is presented in the form where the boundary-free and boundary-induced
contributions are explicitly separated. The properties of the latter are
investigated in various asymptotic regions of the parameters. A similar
investigation for the interior region is presented in section \ref{sec:Int}.
We also discuss the FC for the second type of the boundary condition
differing from the previous one by the sign of the term with the normal to
the boundary. The parity and time-reversal invariant fermionic models in
(2+1)-dimensions can be constructed by combining two spinor fields realizing
two inequivalent representations of the Clifford algebra. The FC in this
class of models and corresponding applications to graphitic cones are
discussed in section \ref{sec:PT}. The main results of the paper are
summarized in section \ref{sec:Conc}. In appendix \ref{sec:appA} we describe
the evaluation of the FC for a field with zero chemical potential and show
that, though the evaluation procedure is different, the final result can be
obtained from the corresponding expression for the nonzero chemical
potential taking the zero chemical potential limit. In appendix \ref%
{sec:AppB} we consider the zero temperature limit and show that both the
representations for the FC give the same result.

\section{Problem setup and the fermionic modes}

\label{sec:modes}

In this section we describe the bulk and boundary geometries for the problem
under consideration and present complete set of fermionic modes outside and
inside a circular boundary. The metric tensor for the background geometry is
given by the (2+1)-dimensional line element
\begin{equation}
ds^{2}=g_{\mu \nu }dx^{\mu }dx^{\nu }=dt^{2}-dr^{2}-r^{2}d\phi ^{2}\ ,
\label{linel}
\end{equation}%
with the spatial coordinates defined in the ranges $r\geqslant 0$ and $%
0\leqslant \phi \leqslant \phi _{0}$. For $\phi _{0}=2\pi $ this corresponds
to the standard (2+1)-dimensional Minkwoski spacetime. In the case $\phi
_{0}<2\pi $ one has a planar angle deficit $2\pi -\phi _{0}$ and the spatial
geometry presents a cone with the apex at $r=0$. In what follows, in
addition to $\phi _{0}$ we will also use the parameter $q=2\pi /\phi _{0}$,
assuming that $q\geqslant 1$. We will consider the case of two-components
spinor field $\psi (x)$ realizing the irreducible representation of the
Clifford algebra. Also we assume the presence of an external electromagnetic
field with the vector potential $A_{\mu }$. The field operator obeys the
Dirac equation
\begin{equation}
\left( i\gamma ^{\mu }D_{\mu }-sm\right) \psi (x)=0,\ D_{\mu }=\partial
_{\mu }+\Gamma _{\mu }+ieA_{\mu },  \label{Dirac}
\end{equation}%
where $\Gamma _{\mu }$ is the spin connection and $e$ the charge of the
field quanta. Here, $s=+1$ and $s=-1$ correspond to two inequivalent
irreducible representations of the Clifford algebra in $(2+1)$-dimensions
(see the discussion in section \ref{sec:PT} below). With these
representations, the mass term violates the parity and time-reversal
invariances \cite{Dese82}. In the coordinates corresponding to (\ref{linel}%
), the gamma matrices can be taken in the representation%
\begin{equation}
\gamma ^{0}=\left(
\begin{array}{cc}
1 & 0 \\
0 & -1%
\end{array}%
\right) ,\quad \gamma ^{l}=\frac{i^{2-l}}{r^{l-1}}\left(
\begin{array}{cc}
0 & e^{-iq\phi } \\
(-1)^{l-1}e^{iq\phi } & 0%
\end{array}%
\right) ,  \label{gamma}
\end{equation}%
with $l=1,2$.

We consider the vector potential of the form $A_{\mu }=(0,0,A)$, where $%
A_{2}=A$ represents the angular component in the coordinates system defined
by $(t,r,\phi )$. For the physical component of the vector potential one has
$A_{\phi }=-A/r$. This corresponds to an infinitely thin magnetic flux $\Phi
=-\phi _{0}A$ located at $r=0$. As it will be seen below, in the expressions
for the FC the parameter $A$ enters in the form of the combination
\begin{equation}
\alpha =eA/q=-e\Phi /(2\pi )\ .  \label{alphan}
\end{equation}%
We decompose it as
\begin{equation}
\alpha =\alpha _{0}+n_{0},\quad |\alpha _{0}|<1/2,  \label{alpha}
\end{equation}%
with $n_{0}$ being an integer. As we will see the FC depends only on the
fractional part $\alpha _{0}$ only.

Now we assume the presence of the circular boundary at $r=a$ on which the
field obeys the MIT bag boundary condition%
\begin{equation}
\left( 1+in_{\mu }\gamma ^{\mu }\right) \psi (x)=0,\;r=a,  \label{BCMIT}
\end{equation}%
with $n_{\mu }$ being the inward pointing unit vector normal to the
boundary. One has $n_{\mu }=\delta _{\mu }^{1}$ and $n_{\mu }=-\delta _{\mu
}^{1}$ in the regions $r\leqslant a$ and $r\geqslant a$, respectively. The
main objective of this paper is to investigate the influence of the boundary
on the FC assuming that the field is in thermal equilibrium at temperature $%
T $. The FC is defined as%
\begin{equation}
\langle \bar{\psi}\psi \rangle =\mathrm{tr\,}[\hat{\rho}\bar{\psi}\psi ],
\label{FC}
\end{equation}
where $\bar{\psi}=\psi ^{\dag }\gamma ^{0}$ is the Dirac adjoint and the
angular brackets denote the ensemble average with the density matrix%
\begin{equation}
\hat{\rho}=\frac{1}{Z}e^{-\beta (\hat{H}-\mu ^{\prime }\hat{Q})},\;\beta =%
\frac{1}{T}.  \label{ro}
\end{equation}
Here $\hat{H}$ is the Hamilton operator, $\hat{Q}$ is a conserved charge
with the related chemical potential $\mu ^{\prime }$ and $Z=\mathrm{tr\,}%
[e^{-\beta (\hat{H}-\mu ^{\prime }\hat{Q})}]$.

Let $\{\psi _{\sigma }^{(+)}(x),\psi _{\sigma }^{(-)}(x)\}$ be a complete
orthonormal set of the positive- and negative-energy solutions of the field
equation (\ref{Dirac}), specified by a set of quantum numbers $\sigma $.
Expanding the field operator $\psi (x)$ in terms of $\psi _{\sigma }^{(\pm
)}(x)$, the FC is decomposed as%
\begin{equation}
\langle \bar{\psi}\psi \rangle =\langle \bar{\psi}\psi \rangle _{\mathrm{vac}%
}+\langle \bar{\psi}\psi \rangle _{T+}+\langle \bar{\psi}\psi \rangle _{T-}.
\label{FCdec}
\end{equation}%
Here,
\begin{equation}
\langle \bar{\psi}\psi \rangle _{\mathrm{vac}}=\sum_{\sigma }\bar{\psi}%
_{\sigma }^{(-)}(x)\psi _{\sigma }^{(-)}(x),  \label{FCvac}
\end{equation}%
is the FC in the vacuum state and $\langle \bar{\psi}\psi \rangle _{T\pm }$
are the contributions from particles (upper sign) and antiparticles (lower
signs). They are given by
\begin{equation}
\langle \bar{\psi}\psi \rangle _{T\pm }=\pm \sum_{\sigma }\frac{\bar{\psi}%
_{\sigma }^{(\pm )}(x)\psi _{\sigma }^{(\pm )}(x)}{e^{\beta (E_{\sigma }\mp
\mu )}+1},  \label{FCpm}
\end{equation}%
where $\mu =e\mu ^{\prime }$ and $\pm E_{\sigma }$ are the energies
corresponding to the modes $\psi _{\sigma }^{(\pm )}(x)$. In (\ref{FCvac})
and (\ref{FCpm}), $\sum_{\sigma }$ includes the summation over the discrete
quantum numbers and the integration over the continuous ones. The modes are
normalized in accordance with the standard normalization condition%
\begin{equation}
\int d^{2}x\,r\psi _{\sigma }^{(\pm )\dag }(x)\psi _{\sigma ^{\prime
}}^{(\pm )}(x)=\delta _{\sigma \sigma ^{\prime }},  \label{NC}
\end{equation}%
where the radial integration goes over the region under consideration. The
part in the FC corresponding to the vacuum expectation value, $\langle \bar{%
\psi}\psi \rangle _{\mathrm{vac}}$, has been investigated in \cite{Bell11}
and here we will be mainly concerned with the finite temperature parts $%
\langle \bar{\psi}\psi \rangle _{T\pm }$. In order to evaluate these parts
we need to specify the mode functions $\psi _{\sigma }^{(\pm )}(x)$.

First let us consider the exterior region, $r\geqslant a$. The corresponding
mode functions are specified by the quantum numbers $(\gamma ,j)$, with $%
0\leqslant \gamma <\infty $, $j=\pm 1/2,\pm 3/2,\ldots $, and have the form
\begin{equation}
\psi _{\sigma }^{(\pm )}(x)=c_{\mathrm{e}}^{(\pm )}e^{iqj\phi \mp iEt}\left(
\begin{array}{c}
g_{\beta _{j},\beta _{j}}^{(\pm )}(\gamma a,\gamma r)e^{-iq\phi /2} \\
\epsilon _{j}\frac{\gamma e^{iq\phi /2}}{\pm E+sm}g_{\beta _{j},\beta
_{j}+\epsilon _{j}}^{(\pm )}(\gamma a,\gamma r)%
\end{array}%
\right) \ ,  \label{psie}
\end{equation}%
where $E=E_{\sigma }=\sqrt{\gamma ^{2}+m^{2}}$, $\epsilon _{j}=1$ for $%
j>-\alpha $ and $\epsilon _{j}=-1$ for $j<-\alpha $,
\begin{equation}
\beta _{j}=q|j+\alpha |-\epsilon _{j}/2.  \label{betj}
\end{equation}%
The function $g_{\beta _{j},\nu }^{(\pm )}(\gamma a,\gamma r)$, with $\nu
=\beta _{j}$ and $\nu =\beta _{j}+\epsilon _{j}$, is expressed in terms of
the Bessel and Neumann functions as:%
\begin{equation}
g_{\beta _{j},\nu }^{(\pm )}(\gamma a,\gamma r)=\bar{Y}_{\beta _{j}}^{(\pm
)}(\gamma a)J_{\nu }(\gamma r)-\bar{J}_{\beta _{j}}^{(\pm )}(\gamma a)Y_{\nu
}(\gamma r)\ .  \label{gbet}
\end{equation}%
Here the notation with the bar is defined as%
\begin{eqnarray}
\bar{F}_{\beta _{j}}^{(\pm )}(z) &=&zF_{\beta _{j}}^{\prime }(z)-\left( \pm
\sqrt{z^{2}+m_{a}^{2}}+sm_{a}+\epsilon _{j}\beta _{j}\right) F_{\beta
_{j}}(z)  \notag \\
&=&-\epsilon _{j}zF_{\beta _{j}+\epsilon _{j}}(z)-\left( \pm \sqrt{%
z^{2}+m_{a}^{2}}+sm_{a}\right) F_{\beta _{j}}(z),  \label{Fbar}
\end{eqnarray}%
and $m_{a}=ma$. The relative coefficient of the linear combination of the
Bessel and Neumann functions in (\ref{gbet}) is determined by the boundary
condition (\ref{BCMIT}). The normalization coefficient $c_{\mathrm{e}}^{(\pm
)}$ is obtained from the condition (\ref{NC}) with the radial integration
over $[a,\infty )$ and with $\delta _{\sigma \sigma ^{\prime }}=\delta
(\gamma -\gamma ^{\prime })\delta _{jj^{\prime }}$. It is given by
\begin{equation}
|c_{\mathrm{e}}^{(\pm )}|^{2}=\frac{\gamma }{2\phi _{0}E}\frac{E\pm sm}{\bar{%
J}_{\beta _{j}}^{(\pm )2}(\gamma a)+\bar{Y}_{\beta _{j}}^{(\pm )2}(\gamma a)}%
\ .  \label{norm}
\end{equation}

In the interior region, $r\leqslant a$, the mode functions are given as%
\begin{equation}
\psi _{\sigma }^{(\pm )}(x)=c_{\mathrm{i}}^{(\pm )}e^{iqj\phi \mp iEt}\left(
\begin{array}{c}
J_{\beta _{j}}(\gamma r)e^{-iq\phi /2} \\
\epsilon _{j}\frac{\gamma e^{iq\phi /2}}{\pm E+sm}J_{\beta _{j}+\epsilon
_{j}}(\gamma r)%
\end{array}%
\right) \ .  \label{psii}
\end{equation}%
From the boundary condition (\ref{BCMIT}) it follows that the eigenvalues of
$\gamma $ are solutions of the equation%
\begin{equation}
\tilde{J}_{\beta _{j}}^{(\pm )}(\gamma a)=0,  \label{modesi}
\end{equation}%
where the notation with tilde for the cylinder functions is defined as
\begin{eqnarray}
\tilde{F}_{\beta _{j}}^{(\pm )}(z) &=&zF_{\beta _{j}}^{\prime }(z)+\left(
\pm \sqrt{z^{2}+m_{a}^{2}}+sm_{a}-\epsilon _{j}\beta _{j}\right) F_{\beta
_{j}}(z)  \notag \\
&=&-\epsilon _{j}zF_{\beta _{j}+\epsilon _{j}}(z)+\left( \pm \sqrt{%
z^{2}+m_{a}^{2}}+sm_{a}\right) F_{\beta _{j}}(z).  \label{Ftilde}
\end{eqnarray}%
We denote the positive roots of the equation (\ref{modesi}) by $\gamma
a=\gamma _{j,l}^{(\pm )}$, $l=1,2,\ldots $. It can be seen that the modes
for the positive energy solution with $j>-\alpha $ coincide with the modes
for negative energy solution with $j<-\alpha $ if we replace $\alpha
\rightarrow -\alpha $ (particles replaced by antiparticles).

The normalization constant $c_{\mathrm{i}}^{(\pm )}$ is determined from (\ref%
{NC}) with the radial integration over $[0,a]$ and $\delta _{\sigma \sigma
^{\prime }}=\delta _{ll^{\prime }}\delta _{jj^{\prime }}$:%
\begin{equation}
c_{\mathrm{i}}^{(\pm )2}=\frac{\gamma }{2\phi _{0}a}\frac{E\pm sm}{E}%
T_{\beta _{j}}(\gamma a),  \label{ci}
\end{equation}%
where we have defined%
\begin{equation}
T_{\beta _{j}}^{(\pm )}(z)=\frac{zJ_{\beta _{j}}^{-2}(z)}{z^{2}+\left(
sm_{a}-\epsilon _{j}\beta _{j}\right) \left( sm_{a}\pm aE\right) \mp \frac{%
z^{2}}{2aE}},\;z=\gamma a.  \label{Tpm}
\end{equation}%
with $aE=\sqrt{z^{2}+m_{a}^{2}}$ and $z=\gamma _{j,l}^{(\pm )}$.

We have determined complete sets of fermionic mode functions outside and
inside the circular boundary with the boundary condition (\ref{BCMIT}) on
it. At this point two comments should be made. The radial functions of the
modes are solutions of the Bessel equation. In the exterior region these
functions are uniquely determined by the boundary condition on the circle $%
r=a$. Inside the circular boundary and for $2|\alpha _{0}|\leqslant 1-1/q$
the fermionic modes are uniquely determined by the normalizability condition
and, as the solution of the Bessel equation, the function $J_{\beta
_{j}}(\gamma r)$ must be taken. In the case $2|\alpha _{0}|>1-1/q$ and for
the mode with $j=-\mathrm{sgn}(\alpha _{0})/2$ both the solutions with the
functions $J_{\beta _{j}}(\gamma r)$ and $Y_{\beta _{j}}(\gamma r)$ are
normalizable. The general solution is a linear combination of these
function. One of the coefficients is determined from the normalization
conditions of the modes. In order to determine the second coefficient, a
boundary condition on the cone apex must be specified. Here the situation is
similar to that for the region around an Aharonov-Bohm gauge field. For the
latter problem it is well-known that the theory of von Neumann deficiency
indices leads to a one-parameter family of allowed boundary conditions \cite%
{Sous89} (see also \cite{Site07} for a discussion related to graphene with a
topological defect). The boundary condition for our choice of the modes (\ref%
{psii}) in the case $j=-\mathrm{sgn}(\alpha _{0})/2$ corresponds to the
situation when the bag boundary condition is imposed on the circle $%
r=\varepsilon $ with small $\varepsilon >0$ and then the limit $\varepsilon
\rightarrow 0$ is taken.

The second comment is related to the periodicity condition with respect to
the rotation around the apex. The mode functions (\ref{psie}) and (\ref{psii}%
) are periodic with respect to that rotation: $\psi _{\sigma }^{(\pm
)}(t,r,\phi +\phi _{0})=\psi _{\sigma }^{(\pm )}(t,r,\phi )$. We can
consider a more general quasiperiodicity condition
\begin{equation}
\psi _{\sigma }^{(\pm )}(t,r,\phi +\phi _{0})=e^{2\pi i\chi }\psi _{\sigma
}^{(\pm )}(t,r,\phi ),  \label{QPC}
\end{equation}%
with a constant phase $2\pi \chi $. The corresponding mode functions are
simply obtained from (\ref{psie}) and (\ref{psii}) by the replacement $%
j\rightarrow j+\chi $. The physical results will depend on $A$ and $\chi $
in the form of the combination $\tilde{\alpha }=\alpha +\chi =eA/q+\chi $.
Though the separate terms $\alpha $ and $\chi $ are gauge dependent, the
combination $\tilde{\alpha }$ is gauge invariant. The results for a field
obeying the quasiperiodicity condition with the phase $2\pi \chi $ are
obtained from those given below by the replacement $\alpha \rightarrow $ $%
\tilde{\alpha }$.

\section{FC in the exterior region}

\label{sec:Ext}

Having specified the mode functions, we start our investigation for the FC
in the exterior region. In that region, the FC\ in the vacuum state is
decompsed as \cite{Bell11}%
\begin{equation}
\langle \bar{\psi}\psi \rangle _{\mathrm{vac}}=\langle \bar{\psi}\psi
\rangle _{\mathrm{vac}}^{(0)}+\langle \bar{\psi}\psi \rangle _{\mathrm{vac}%
}^{(b)},  \label{FCvac2}
\end{equation}%
where the FC\ for the vacuum state in the boundary-free geometry is given by
the expression%
\begin{eqnarray}
&&\langle \bar{\psi}\psi \rangle _{\mathrm{vac}}^{(0)}=-\frac{sm}{2\pi r}%
\Big\{\sum_{l=1}^{[q/2]}(-1)^{l}\frac{\cot (\pi l/q)}{e^{2mr\sin (\pi l/q)}}%
\cos (2\pi l\alpha _{0})  \notag \\
&&\qquad -\frac{q}{\pi }\int_{0}^{\infty }dy\frac{e^{-2mr\cosh y}}{\cosh y}%
\frac{f_{1}(q,\alpha _{0},y)}{\cosh (2qy)-\cos (q\pi )}\Big\},  \label{FC0}
\end{eqnarray}%
with $[q/2]$ being the integer part of $q/2$ and%
\begin{equation}
f_{1}(q,\alpha _{0},y)=-\sinh y\sum_{\delta =\pm 1}\cos (q\pi (1/2-\delta
\alpha _{0}))\sinh ((1+2\delta \alpha _{0})qy).  \label{f1}
\end{equation}%
For the boundary-induced contribution in the vacuum state one has
\begin{eqnarray}
\langle \bar{\psi}\psi \rangle _{\mathrm{vac}}^{(b)} &=&-\frac{1}{\pi \phi
_{0}}\sum_{j}\int_{m}^{\infty }dx\,x\left\{ \mathrm{Im}\left[ \frac{\bar{I}%
_{\beta _{j}}(xa)}{\bar{K}_{\beta _{j}}(xa)}\right] \left[ K_{\beta
_{j}}^{2}(xr)+K_{\beta _{j}+\epsilon _{j}}^{2}(xr)\right] \right.  \notag \\
&&+\left. sm\mathrm{Re}\left[ \frac{\bar{I}_{\beta _{j}}(xa)}{\bar{K}_{\beta
_{j}}(xa)}\right] \frac{K_{\beta _{j}+\epsilon _{j}}^{2}(xr)-K_{\beta
_{j}}^{2}(xr)}{\sqrt{x^{2}-m^{2}}}\right\} ,  \label{FCbvac}
\end{eqnarray}%
where for the modified Bessel functions we use the notation
\begin{equation}
\bar{F}_{\beta _{j}}(u)=uF_{\beta _{j}}^{\prime }(u)-\left( i\sqrt{%
u^{2}-m_{a}^{2}}+sm_{a}+\epsilon _{j}\beta _{j}\right) F_{\beta _{j}}(u).
\label{FbarIK}
\end{equation}%
Note that for $1\leq q<2$ the first term in figure braces of (\ref{FC0}) is
absent. Here we are interested in the finite temperature contributions.

Taking into account (\ref{FCpm}) and (\ref{psie}), after some intermediate
steps we get
\begin{eqnarray}
\langle \bar{\psi}\psi \rangle _{T\pm } &=&\pm \frac{1}{2\phi _{0}}%
\sum_{j}\int_{0}^{\infty }d\gamma \frac{\gamma /E}{e^{\beta (E\mp \mu )}+1}
\notag \\
&&\times \frac{(E\pm sm)g_{\beta _{j},\beta _{j}}^{(\pm )2}(\gamma a,\gamma
r)-(E\mp sm)g_{\beta _{j},\beta _{j}+\epsilon _{j}}^{(\pm )2}(\gamma
a,\gamma r)}{\bar{J}_{\beta _{j}}^{(\pm )2}(\gamma a)+\bar{Y}_{\beta
_{j}}^{(\pm )2}(\gamma a)}.  \label{FCpme}
\end{eqnarray}%
In order to find an explicit expression for the boundary-induced part, we
subtract form (\ref{FCpme}) the corresponding boundary-free term $\langle
\bar{\psi}\psi \rangle _{T\pm }^{(0)}$. The expression for the latter is
obtained from (\ref{FCpme}) by the replacements $g_{\beta _{j},\nu }^{(\pm
)2}(x,y)/[\bar{J}_{\beta _{j}}^{(\pm )2}(x)+\bar{Y}_{\beta _{j}}^{(\pm
)2}(x)]\rightarrow J_{\nu }^{2}(y)$ with $\nu =\beta _{j}$ and $\nu =\beta
_{j}+\epsilon _{j}$ (see \cite{Bell16T}). For the evaluation of the
boundary-induced part, we use the identity
\begin{equation}
\frac{g_{\beta _{j},\nu }^{(\pm )2}(x,y)}{\bar{J}_{\beta _{j}}^{(\pm )2}(x)+%
\bar{Y}_{\beta _{j}}^{(\pm )2}(x)}-J_{\nu }^{2}(y)=-\frac{1}{2}\sum_{l=1,2}%
\frac{\bar{J}_{\beta _{j}}^{(\pm )}(x)}{\bar{H}_{\beta _{j}}^{(l,\pm )}(x)}%
H_{\nu }^{(l)2}(y),  \label{ident1}
\end{equation}%
valid for both $\nu =\beta _{j},\beta _{j}+\epsilon _{j}$, and with $H_{\nu
}^{(l)}(x)$ being the Hankel functions. In this way, for the
boundary-induced parts%
\begin{equation}
\langle \bar{\psi}\psi \rangle _{T\pm }^{(b)}=\langle \bar{\psi}\psi \rangle
_{T\pm }-\langle \bar{\psi}\psi \rangle _{T\pm }^{(0)}  \label{FCpmb}
\end{equation}%
we get
\begin{eqnarray}
\langle \bar{\psi}\psi \rangle _{T\lambda }^{(b)} &=&-\lambda \frac{1}{4\phi
_{0}}\sum_{j}\sum_{l=1,2}\int_{0}^{\infty }d\gamma \frac{\gamma }{E}\frac{%
\bar{J}_{\beta _{j}}^{(\lambda )}(\gamma a)}{\bar{H}_{\beta
_{j}}^{(l,\lambda )}(\gamma a)}  \notag \\
&&\times \frac{(E+\lambda sm)H_{\beta _{j}}^{(l)2}(\gamma r)-(E-\lambda
sm)H_{\beta _{j}+\epsilon _{j}}^{(l)2}(\gamma r)}{e^{\beta (E-\lambda \mu
)}+1}\ ,  \label{FCpmb1}
\end{eqnarray}%
with $\lambda =\pm $. For the further transformation of (\ref{FCpmb1}) we
will assume that $\mu \neq 0$. The case $\mu =0$ will be considered in
appendix \ref{sec:appA}.

The integrand in (\ref{FCpmb1}) has simple poles for
\begin{equation}
E=E_{n}^{(\lambda )}\equiv \lambda \mu +i\pi \left( 2n+1\right) T,
\label{En}
\end{equation}%
with $n=0,\pm 1,\pm 2,\ldots $. One has $n=0,1,2,\ldots $ for the poles in
the upper half-plane and $n=\ldots ,-2,-1$ in the lower half-plane. For the
values of $\gamma =\gamma _{n}^{(\lambda )}$ corresponding to the poles (\ref%
{En}) we get
\begin{equation}
\gamma _{n}^{(\lambda )2}=\left[ \lambda \mu +i\pi \left( 2n+1\right) T%
\right] ^{2}-m^{2},  \label{gamn}
\end{equation}%
where, again, $n=0,1,2,\ldots $ ($n=\ldots ,-2,-1$) for the poles in the
upper (lower) half-plane. Note that for the poles in the upper and lower
half-planes one has the relations%
\begin{equation}
E_{n}^{(\lambda )}=E_{-n-1}^{(\lambda )\ast },\;\gamma _{n}^{(\lambda
)}=\gamma _{-n-1}^{(\lambda )\ast },\;n=\ldots ,-2,-1,  \label{RelPole}
\end{equation}
where the star stands for the complex conjugate.

For the transformation of (\ref{FCpmb1}) we rotate the integration contour
in the complex plane $\gamma $ by the angle $\pi /2$ for the term with $l=1$
and by the angle $-\pi /2$ for the term with $l=2$. For $\lambda \mu <0$ the
thermal factor $1/[e^{\beta (E-\lambda \mu )}+1]$ has no poles in the right
half-plane and the integral is transformed to the integrals over the
imaginary axis. In the case $\lambda \mu >0$, in addition to the latter
integrals the residue terms from the poles (\ref{gamn}) should be added. In
the integral over the positive imaginary semiaxis we introduce the modified
Bessel functions $I_{\nu }(z)$ and $K_{\nu }(z)$ by using the relations%
\begin{equation}
\bar{J}_{\beta _{j}}^{(\lambda )}(e^{\pi i/2}z)=e^{i\pi \beta _{j}/2}\bar{I}%
_{\beta _{j}}^{(\lambda )}(z),\;\bar{H}_{\beta _{j}}^{(1,\lambda )}(e^{\pi
i/2}z)=\frac{2}{\pi i}e^{-i\pi \beta _{j}/2}\bar{K}_{\beta _{j}}^{(\lambda
)}(z),  \label{Jb}
\end{equation}%
where for the modified Bessel functions we use the notation%
\begin{equation}
\bar{F}_{\beta _{j}}^{(\lambda )}(z)=zF_{\beta _{j}}^{\prime }(z)-\left(
\lambda \sqrt{\left( e^{\pi i/2}z\right) ^{2}+m_{a}^{2}}+sm_{a}+\epsilon
_{j}\beta _{j}\right) F_{\beta _{j}}(z),  \label{FbIK}
\end{equation}%
with $F=I,K$. For the functions in the integral over the negative imaginary
semiaxis one has $\bar{J}_{\beta _{j}}^{(\lambda )}(e^{-\pi i/2}z)=e^{i\pi
\beta _{j}/2}\bar{I}_{\beta _{j}}^{(\lambda )\ast }(z)$ and $\bar{H}_{\beta
_{j}}^{(2,\lambda )}(e^{-\pi i/2}z)=-\frac{2}{\pi i}e^{i\pi \beta _{j}/2}%
\bar{K}_{\beta _{j}}^{(\lambda )\ast }(z)$. Note that for $z\geqslant 0$ the
square root is understood as%
\begin{equation}
\sqrt{\left( e^{\pm \pi i/2}z\right) ^{2}+m_{a}^{2}}=\left\{
\begin{array}{cc}
\sqrt{m_{a}^{2}-z^{2}}, & z<m_{a}, \\
e^{\pm \pi i/2}\sqrt{z^{2}-m_{a}^{2}}, & z>m_{a}.%
\end{array}%
\right.  \label{root}
\end{equation}%
From here it follows that $\bar{F}_{\beta _{j}}^{(\lambda )\ast }(z)=\bar{F}%
_{\beta _{j}}^{(\lambda )}(z)$ for $z<m_{a}$. By using this relation we can
see that the integrals over the intervals $(0,im_{a})$ and $(0,-im_{a})$
cancel each other. For $\lambda \mu >0$, the contributions to $\langle \bar{%
\psi}\psi \rangle _{b,\lambda }^{(T)}$ from the residue terms at the poles
in the upper and lower half-planes are combined as
\begin{equation}
-\lambda \frac{\pi }{\phi _{0}}\theta \left( \lambda \mu \right)
T\sum_{j}\sum_{n=0}^{\infty }\mathrm{Im}\left\{ \frac{\bar{J}_{\beta
_{j}}^{(\lambda )}(\gamma _{n}^{(\lambda )}a)}{\bar{H}_{\beta
_{j}}^{(1,\lambda )}(\gamma _{n}^{(\lambda )}a)}\left[ (E_{n}^{(\lambda
)}+\lambda sm)H_{\beta _{j}}^{(1)2}(\gamma _{n}^{(\lambda
)}r)-(E_{n}^{(\lambda )}-\lambda sm)H_{\beta _{j}+\epsilon
_{j}}^{(1)2}(\gamma _{n}^{(\lambda )}r)\right] \right\} ,  \label{PoleCont}
\end{equation}%
where $\theta \left( x\right) $ is the Heaviside step function and for the
poles in the lower half-plane we have used the relations (\ref{RelPole}). We
find it convenient to introduce in (\ref{PoleCont}) a new quantity $%
u_{n}^{(\lambda )}$ in accordance with $\gamma _{n}^{(\lambda
)}=iu_{n}^{(\lambda )}$, $\mathrm{Re\,}u_{n}^{(\lambda )}>0$,
\begin{equation}
u_{n}^{(\lambda )}=\{\left[ \pi \left( 2n+1\right) T-i\lambda \mu \right]
^{2}+m^{2}\}^{1/2}.  \label{unlam}
\end{equation}%
Note that $u_{n}^{(-)}=u_{n}^{(+)\ast }$.

After the transformations described above, the boundary-induced contribution
in the thermal part of the FC is presented as%
\begin{eqnarray}
\langle \bar{\psi}\psi \rangle _{T\lambda }^{(b)} &=&\lambda \frac{1}{\pi
\phi _{0}}\sum_{j}\int_{m}^{\infty }dx\,x\mathrm{Im}\left\{ \frac{\bar{I}%
_{\beta _{j}}^{(\lambda )}(xa)}{\bar{K}_{\beta _{j}}^{(\lambda )}(xa)}\frac{1%
}{e^{\beta (i\sqrt{x^{2}-m^{2}}-\lambda \mu )}+1}\right.  \notag \\
&&\times \left. \left[ (1-\frac{i\lambda sm}{\sqrt{x^{2}-m^{2}}})K_{\beta
_{j}}^{2}(xr)+(1+\frac{i\lambda sm}{\sqrt{x^{2}-m^{2}}})K_{\beta
_{j}+\epsilon _{j}}^{2}(xr)\right] \right\}  \notag \\
&&-\lambda \frac{2}{\phi _{0}}\theta \left( \lambda \mu \right)
T\sum_{j}\sum_{n=0}^{\infty }\mathrm{Im}\left\{ \frac{\bar{I}_{\beta
_{j}}^{(\lambda )}(u_{n}^{(\lambda )}a)}{\bar{K}_{\beta _{j}}^{(\lambda
)}(u_{n}^{(\lambda )}a)}\left[ (\pi \left( 2n+1\right) T-i\lambda \left( \mu
+sm\right) )K_{\beta _{j}}^{2}(u_{n}^{(\lambda )}r)\right. \right.  \notag \\
&&\left. \left. +(\pi \left( 2n+1\right) T-i\lambda \left( \mu -sm\right)
)K_{\beta _{j}+\epsilon _{j}}^{2}(u_{n}^{(\lambda )}r)\right] \right\} .
\label{FCb2}
\end{eqnarray}%
By taking into account the relations $\bar{I}_{\beta _{j}}^{(-)}(xa)=\bar{I}%
_{\beta _{j}}^{(+)\ast }(xa)$, $\bar{K}_{\beta _{j}}^{(-)}(xa)=\bar{K}%
_{\beta _{j}}^{(+)\ast }(xa)$, we can see that for $\lambda =-$ the
expressions under the sign of the summation over $n$ in (\ref{FCb2}) differs
from that for $\lambda =+$ by the sign. As a consequence, the expression (%
\ref{FCb2}) is transformed to%
\begin{eqnarray}
\langle \bar{\psi}\psi \rangle _{T\lambda }^{(b)} &=&\frac{1}{\pi \phi _{0}}%
\sum_{j}\int_{m}^{\infty }dx\,x\mathrm{Im}\left\{ \frac{\bar{I}_{\beta
_{j}}(xa)}{\bar{K}_{\beta _{j}}(xa)}\frac{1}{e^{\lambda \beta (i\sqrt{%
x^{2}-m^{2}}-\mu )}+1}\right.  \notag \\
&&\times \left. \left[ \left( 1-\frac{ism}{\sqrt{x^{2}-m^{2}}}\right)
K_{\beta _{j}}^{2}(xr)+\left( 1+\frac{ism}{\sqrt{x^{2}-m^{2}}}\right)
K_{\beta _{j}+\epsilon _{j}}^{2}(xr)\right] \right\}  \notag \\
&&-\frac{2}{\phi _{0}}\theta \left( \lambda \mu \right)
T\sum_{j}\sum_{n=0}^{\infty }\mathrm{Im}\left\{ \frac{\bar{I}_{\beta
_{j}}(u_{n}a)}{\bar{K}_{\beta _{j}}(u_{n}a)}\left[ (\pi \left( 2n+1\right)
T-i\left( \mu +sm\right) )K_{\beta _{j}}^{2}(u_{n}r)\right. \right.  \notag
\\
&&\left. \left. +(\pi \left( 2n+1\right) T-i\left( \mu -sm\right) )K_{\beta
_{j}+\epsilon _{j}}^{2}(u_{n}r)\right] \right\} .  \label{FCb3}
\end{eqnarray}%
Here
\begin{equation}
u_{n}=\{\left[ \pi \left( 2n+1\right) T-i\mu \right] ^{2}+m^{2}\}^{1/2},
\label{un}
\end{equation}%
and for the modified Bessel functions we use the notation $\bar{F}_{\beta
_{j}}(z)=\bar{F}_{\beta _{j}}^{(+)}(z)$, defined by (\ref{FbarIK}) with $%
F=I,K$. The expressions (\ref{FCb3}) with $\lambda =+$ and $\lambda =-$
present the contributions to the boundary-induced FC coming from particles
and antiparticles.

Combining the contribution from the separate terms for $\lambda =+$ and $%
\lambda =-$, we can see that in evaluating the boundary-induced part $%
\langle \bar{\psi}\psi \rangle _{T}^{(b)}=\sum_{\lambda =\pm }\langle \bar{%
\psi}\psi \rangle _{T\lambda }^{(b}$ the sum of the first terms in the
right-hand side of (\ref{FCb3}) is equal to $-\langle \bar{\psi}\psi \rangle
_{\mathrm{vac}}^{(b)}$ with $\langle \bar{\psi}\psi \rangle _{\mathrm{vac}%
}^{(b)}$ from (\ref{FCbvac}). Hence, the boundary-induced contribution at
temperature $T$, given by%
\begin{equation}
\langle \bar{\psi}\psi \rangle ^{(b)}=\langle \bar{\psi}\psi \rangle _{%
\mathrm{vac}}^{(b)}+\langle \bar{\psi}\psi \rangle _{T}^{(b)},  \label{FCbe}
\end{equation}%
is presented in the form
\begin{eqnarray}
\langle \bar{\psi}\psi \rangle ^{(b)} &=&-\frac{2T}{\phi _{0}}%
\sum_{j}\sum_{n=0}^{\infty }\mathrm{Im}\left\{ \frac{\bar{I}_{\beta
_{j}}(u_{n}a)}{\bar{K}_{\beta _{j}}(u_{n}a)}\left[ (\pi \left( 2n+1\right)
T-i\left( \mu +sm\right) )K_{\beta _{j}}^{2}(u_{n}r)\right. \right.  \notag
\\
&&\left. \left. +(\pi \left( 2n+1\right) T-i\left( \mu -sm\right) )K_{\beta
_{j}+\epsilon _{j}}^{2}(u_{n}r)\right] \right\} .  \label{FCb4}
\end{eqnarray}%
The ratio under the imaginary part in this expression can be written in the
form%
\begin{equation}
\frac{\bar{I}_{\beta _{j}}(z)}{\bar{K}_{\beta _{j}}(z)}=\frac{W_{\beta
_{j},\beta _{j}+\epsilon _{j}}^{(-)}(z)+\left[ i\pi \left( 2n+1\right) T+\mu %
\right] a/z}{z[K_{\beta _{j}+\epsilon _{j}}^{2}(z)+K_{\beta
_{j}}^{2}(z)]+2sm_{a}K_{\beta _{j}}(z)K_{\beta _{j}+\epsilon _{j}}(z)},
\label{IKrat}
\end{equation}%
with the notation (the notation with the $+$ sign will appear in the
expression for the FC in the interior region)%
\begin{eqnarray}
W_{\beta _{j},\beta _{j}+\epsilon _{j}}^{(\pm )}(z) &=&z\left[ I_{\beta
_{j}}(z)K_{\beta _{j}}(z)-I_{\beta _{j}+\epsilon _{j}}(z)K_{\beta
_{j}+\epsilon _{j}}(z)\right]  \notag \\
&&\pm sm_{a}\left[ I_{\beta _{j}+\epsilon _{j}}(z)K_{\beta _{j}}(z)-I_{\beta
_{j}}(z)K_{\beta _{j}+\epsilon _{j}}(z)\right] ,  \label{Wpm}
\end{eqnarray}%
and with $z=u_{n}a$.

For the total FC one has%
\begin{equation}
\langle \bar{\psi}\psi \rangle =\langle \bar{\psi}\psi \rangle
^{(0)}+\langle \bar{\psi}\psi \rangle ^{(b)},  \label{FCdec2}
\end{equation}%
where $\langle \bar{\psi}\psi \rangle ^{(0)}$ is the FC at temperature $T$
in the absence of the boundary \cite{Bell16T}. The notation with bar in (\ref%
{FCb4}) can also be presented in the form%
\begin{equation}
\bar{F}_{\beta _{j}}(z)=\delta _{F}zF_{\beta _{j}+\epsilon _{j}}(z)-\left( i%
\sqrt{z^{2}-m_{a}^{2}}+sm_{a}\right) F_{\beta _{j}}(z),  \label{Fbar4}
\end{equation}%
where $\delta _{I}=-\delta _{K}=1$. Under the replacements $\alpha
\rightarrow -\alpha $, $j\rightarrow -j$ one has $\beta _{j}\rightleftarrows
\beta _{j}+\epsilon _{j}$. By using this and the representation (\ref{Fbar4}%
), it can be seen that under the same replacements we get%
\begin{equation}
\frac{\bar{I}_{\beta _{j}}(u_{n}a)}{\bar{K}_{\beta _{j}}(u_{n}a)}\rightarrow
-\left[ \frac{\bar{I}_{\beta _{j}}(u_{n}^{\ast }a)}{\bar{K}_{\beta
_{j}}(u_{n}^{\ast }a)}\right] ^{\ast }.  \label{IKrepl}
\end{equation}%
Now, by taking into account the relation $u_{n}^{\ast }(\mu )=u_{n}(-\mu )$,
one can show that $\langle \bar{\psi}\psi \rangle ^{(b)}$ is an even
function under the simultaneous replacements $\alpha \rightarrow -\alpha $, $%
\mu \rightarrow -\mu $.

In \cite{Bell16T}, the boundary-free contribution is presented in the form%
\begin{equation}
\langle \bar{\psi}\psi \rangle ^{(0)}=\langle \bar{\psi}\psi \rangle _{%
\mathrm{M}}^{(0)}+\langle \bar{\psi}\psi \rangle _{\mathrm{t}}^{(0)},
\label{FC0n}
\end{equation}
where
\begin{equation}
\langle \bar{\psi}\psi \rangle _{\mathrm{M}}^{(0)}=\frac{smT}{2\pi }\left[
\ln \left( 1+e^{-(m-\mu )/T}\right) +\ln \left( 1+e^{-(m+\mu )/T}\right) %
\right] ,  \label{FC0M}
\end{equation}%
is the FC in (2+1)-dimensional Minkowski spacetime (the magnetic flux and
the planar angle deficit are absent, $q=1$, $\alpha =0$) and $\langle \bar{%
\psi}\psi \rangle _{\mathrm{t}}^{(0)}$ is the topological part induced by
the conical geometry and by the magnetic flux. The latter is given by the
expression \cite{Bell16T}%
\begin{eqnarray}
\langle \bar{\psi}\psi \rangle _{\mathrm{t}}^{(0)} &=&-\frac{2mT}{\pi }%
\sum_{n=-\infty }^{\infty }\left\{ s\sum_{l=1}^{[q/2]}(-1)^{l}c_{l}\cos
(2\pi l\alpha _{0})K_{0}\left( 2rs_{l}u_{n}\right) \right.  \notag \\
&&-\frac{sq}{\pi }\int_{0}^{\infty }dy\,\frac{f_{1}(q,\alpha
_{0},y)K_{0}\left( 2ru_{n}\cosh y\right) }{\cosh (2qy)-\cos (q\pi )}  \notag
\\
&&+\frac{\mu +i\pi (2n+1)T}{m}\left[ \sum_{l=1}^{[q/2]}(-1)^{l}s_{l}\sin
(2\pi l\alpha _{0})K_{0}\left( 2rs_{l}u_{n}\right) \right.  \notag \\
&&\left. \left. -\frac{q}{\pi }\int_{0}^{\infty }dy\,\frac{f_{2}(q,\alpha
_{0},y)K_{0}\left( 2ru_{n}\cosh y\right) }{\cosh (2qy)-\cos (q\pi )}\right]
\right\} ,  \label{FC03}
\end{eqnarray}%
where $c_{l}=\cos (\pi l/q)$, $s_{l}=\sin (\pi l/q)$, and%
\begin{equation}
f_{2}(q,\alpha _{0},y)=\cosh y\sum_{\delta =\pm 1}\delta \cos (q\pi
(1/2-\delta \alpha _{0}))\cosh ((1+2\delta \alpha _{0})qy).  \label{f2}
\end{equation}%
The representation (\ref{FC03}) is well adapted for the investigation of
high-temperature asymptotic. An alternative representation, convenient in
the low-temperature limit, is provided in Ref. \cite{Bell16T}.

In the case of zero chemical potential, $\mu =0$, the poles of the integrand
in (\ref{FCpmb1}) are located on the imaginary axis and the procedure for
the transformation is different from that we have described above. This case
is considered in appendix \ref{sec:appA}, where it has been shown that the
final result is obtained from (\ref{FCb4}) in the limit $\mu \rightarrow 0$.
The corresponding expression can also be presented in the form%
\begin{eqnarray}
\langle \bar{\psi}\psi \rangle ^{(b)} &=&-\frac{2T}{\phi _{0}}%
\sum_{j}\sum_{n=0}^{\infty }\left\{ \pi \left( 2n+1\right) T\mathrm{Im}\left[
\frac{\bar{I}_{\beta _{j}}(u_{0n}a)}{\bar{K}_{\beta _{j}}(u_{0n}a)}\right] %
\left[ K_{\beta _{j}}^{2}(u_{0n}r)+K_{\beta _{j}+\epsilon _{j}}^{2}(u_{0n}r)%
\right] \right.  \notag \\
&&\left. -sm\mathrm{Re}\left[ \frac{\bar{I}_{\beta _{j}}(u_{0n}a)}{\bar{K}%
_{\beta _{j}}(u_{0n}a)}\right] \left[ K_{\beta _{j}}^{2}(u_{0n}r)-K_{\beta
_{j}+\epsilon _{j}}^{2}(u_{0n}r)\right] \right\} ,  \label{FCbtotmu0}
\end{eqnarray}%
where $u_{0n}$ is defined by%
\begin{equation}
u_{0n}=\sqrt{\left[ \pi \left( 2n+1\right) T\right] ^{2}+m^{2}}.  \label{u0n}
\end{equation}%
The boundary-induced FC (\ref{FCbtotmu0}) is an even function of $\alpha $.
The ratio under the imaginary and real parts is presented in the form (\ref%
{IKrat}), where now $z=u_{0n}r$ is real and the imaginary and real parts are
easily separated. The expression (\ref{FCbtotmu0}) is further simplified for
a massless field:%
\begin{equation}
\langle \bar{\psi}\psi \rangle ^{(b)}=-\frac{2T}{\phi _{0}a}%
\sum_{j}\sum_{n=0}^{\infty }\left. \frac{K_{\beta _{j}}^{2}(ur)+K_{\beta
_{j}+\epsilon _{j}}^{2}(ur)}{K_{\beta _{j}}^{2}(ua)+K_{\beta _{j}+\epsilon
_{j}}^{2}(ua)}\right\vert _{u=\pi \left( 2n+1\right) T}.  \label{FCbm0}
\end{equation}%
Of course, in this case the FC does not depend on the parameter $s$. The
corresponding boundary-free part vanishes, $\langle \bar{\psi}\psi \rangle
^{(0)}=0$ (see \cite{Bell16T}), and the total FC $\langle \bar{\psi}\psi
\rangle =\langle \bar{\psi}\psi \rangle ^{(b)}$ is always negative. It is a
monotonically increasing function of the radial coordinate $r$.

Now we pass to the investigation of the boundary-induced FC in asymptotic
regions for the values of the parameters. For general values of the chemical
potential and the mass, at large distances from the boundary, in (\ref{FCb4}%
) we use the asymptotic expression of the Macdonald function for large
arguments. To the leading order this gives%
\begin{equation}
\langle \bar{\psi}\psi \rangle ^{(b)}\approx -\frac{qT}{r}%
\sum_{j}\sum_{n=0}^{\infty }\mathrm{Im}\left\{ \frac{\bar{I}_{\beta
_{j}}(u_{n}a)}{\bar{K}_{\beta _{j}}(u_{n}a)}\frac{e^{-2u_{n}r}}{u_{n}}\left[
\pi \left( 2n+1\right) T-i\mu \right] \right\} .  \label{FCr}
\end{equation}%
For $T\gtrsim m,|\mu |$ the dominant contribution comes from the term $n=0$
and the boundary-induced contribution is suppressed by the factor $%
e^{-2ru_{0}}$. For $q<2$, a similar suppression takes place for the
topological part $\langle \bar{\psi}\psi \rangle _{\mathrm{t}}^{(0)}$ in the
boundary-free geometry. For $q>2$, the suppression of the latter at large
distances is weaker, by the factor $e^{-2ru_{0}\sin (\pi /q)}$. The
Minkwoskian part (\ref{FC0M}) does not depend on the radial coordinate and
for a massive field it dominates at large distances. For a massless field
with zero chemical potential and for $Tr\gg 1$ one has%
\begin{equation}
\langle \bar{\psi}\psi \rangle ^{(b)}\approx -\frac{2e^{-2\pi Tr}}{\phi
_{0}ar}\sum_{j}\frac{1}{K_{\beta _{j}}^{2}(\pi Ta)+K_{\beta _{j}+\epsilon
_{j}}^{2}(\pi Ta)}.  \label{FCrm0}
\end{equation}%
Hence, the boundary-induced FC is exponentially suppressed at large
distances. Note that at large distances the boundary-induced contribution in
the vacuum FC behaves like $\langle \bar{\psi}\psi \rangle _{\mathrm{vac}%
}^{(b)}\propto e^{-2mr}/r^{2}$, $mr\gg 1$, for a massive field and as $%
\langle \bar{\psi}\psi \rangle _{\mathrm{vac}}^{(b)}\propto
1/r^{q(1-2|\alpha _{0}|)+2}$ in the case of a massless field.

In the high temperature limit, $T\gg m,|\mu |,1/(r-a)$, again, the
contrubution of the $n=0$ term dominates in (\ref{FCb4}) and, similar to (%
\ref{FCr}), we can see that the boundary-induced FC for a given $r$ is
suppressed by the factor $e^{-2\pi Tr}$. For the boundary-free topological
part we have similar behavior, $\langle \bar{\psi}\psi \rangle _{\mathrm{t}%
}^{(0)}\propto $ $e^{-2\pi Tr}$ for $q<2$. In the case $q>2$ one has $%
\langle \bar{\psi}\psi \rangle _{\mathrm{t}}^{(0)}\propto $ $e^{-2\pi Tr\sin
(\pi /q)}$ and the decay is slower. As a consequence, at high temperatures
and for points not too close to the boundary, the total FC is dominated by
the Minkwoskian part that behaves like $\langle \bar{\psi}\psi \rangle _{%
\mathrm{M}}^{(0)}\approx smT\ln 2/(2\pi )$.

The boundary-induced FC (\ref{FCb4}) diverges on the boundary. This kind of
surface divergences in the VEVs of local physical observables are well-known
in quantum field theory with boundaries. They are related to the idealized
boundary conditions on fields acting in the same way for all the modes of
the field. For points near the boundary, assuming that $T(r-a)\ll 1$, the
dominant contribution to the series over $n$ in (\ref{FCb4}) comes from
large $n$ and, to the leading order, we can replace the corresponding
summation by the integration. In Appendix \ref{sec:AppB}, it is shown that,
with this replacement, the corresponding expectation value is obtained for
the vacuum state. Hence, we conclude that for points near the boundary and
for temperatures $T\ll 1/(r-a)$ the finite temperature effects on the FC are
small and the leading term coincides with the vacuum FC. Near the boundary
the latter is dominated by the boundary-induced part and behaves as $\langle
\bar{\psi}\psi \rangle _{\mathrm{vac}}\approx \langle \bar{\psi}\psi \rangle
_{\mathrm{vac}}^{(b)}\approx -1/[8\pi (r-a)^{2}]$. Note that this leading
term does not depend on the planar angle deficit and on the magnetic flux.

It is also of interest to consider the behavior of the boundary-induced FC\
for small values of the radius $a$ and for fixed $r$, assuming that $%
Ta,ma\ll 1$. By using the asymptotic expressions for the modified Bessel
functions for small values of the argument, from (\ref{FCb4}) one can see
that $\langle \bar{\psi}\psi \rangle ^{(b)}\propto a^{q(1-2|\alpha _{0}|)}$
and for $|\alpha _{0}|<1/2$ the boundary-induced contribution tends to zero
in the limit $a\rightarrow 0$. In the special case $|\alpha _{0}|=1/2$ the
part $\langle \bar{\psi}\psi \rangle ^{(b)}$ tends to a finite limiting
value. The case $|\alpha _{0}|=1/2$ is also special for the boundary-free
geometry. For example, the VEVs of the charge and current densities, as
functions of the parameter $\alpha $ from (\ref{alphan}), are discontinous
at the points corresponding to half-odd-integer values of $\alpha $ (see,
for example, Ref. \cite{Beze10}). In accordance with (\ref{alpha}), this
correspond to the case $|\alpha _{0}|=1/2$. A similar feature for the
persistent current in carbon nanotube based rings has been observed in Ref.
\cite{Lin98}. Note that the VEVs of the charge and current densities in the
region $r>a$ vanish for $|\alpha _{0}|=1/2$.

\begin{figure}[tbph]
\begin{center}
\begin{tabular}{cc}
\epsfig{figure=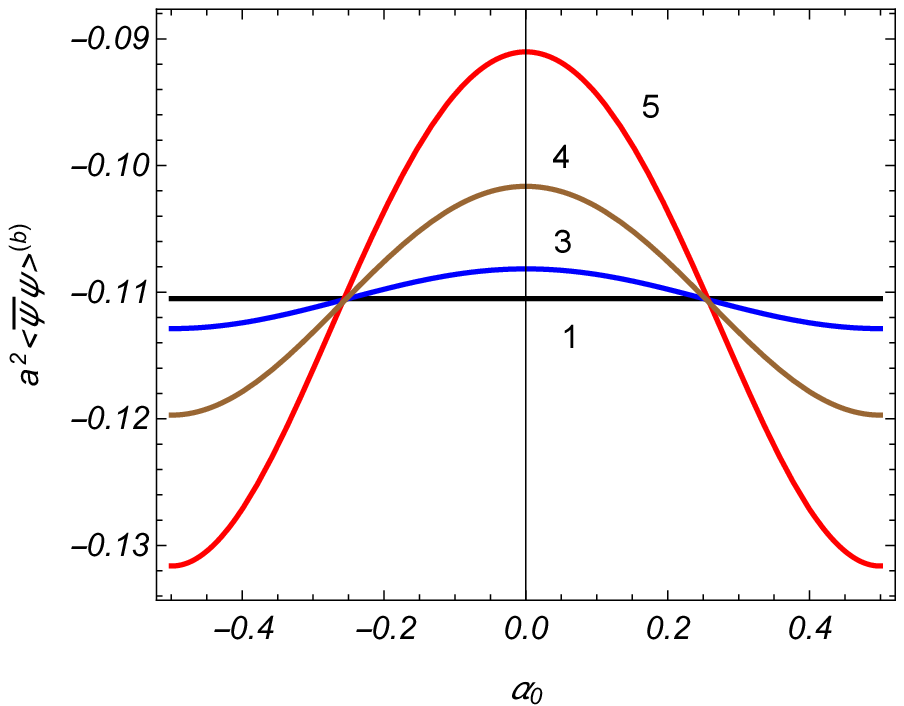,width=7.5cm,height=6.cm} & \quad %
\epsfig{figure=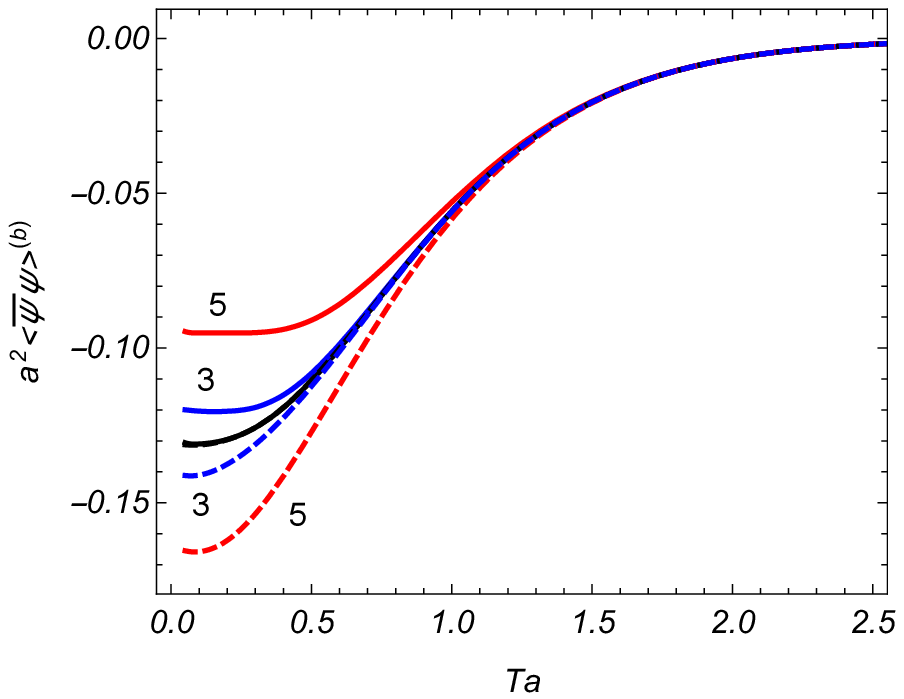,width=7.5cm,height=6.cm}%
\end{tabular}%
\end{center}
\caption{FC in the exterior region for a massless field with a zero chemical
potential versus the parameter $\protect\alpha _{0}$ (left panel) and the
temperature (right panel). The graphs are plotted for $r/a=1.5$ and the
numbers near the curves correspond to the values of $q$. For the left panel
we have taken $Ta=0.5$. The full/dashed curves in the right panel correspond
to $\protect\alpha _{0}=0$/$\protect\alpha _{0}=0.4$.}
\label{fig1}
\end{figure}

The numerical examples for the dependence of the FC on the parameters of the
problem will be given for a simple case of a massless field with a zero
chemical potential (for the effects of the nonzero mass see figure \ref{fig5}
below). In the boundary-free geometry the FC vanishes and the nonzero FC is
induced by the boundary. The left panel in figure \ref{fig1} displays the FC
in the exterior region versus the parameter $\alpha _{0}$ for fixed values
of $r/a=1.5$, $Ta=0.5$. The numbers near the curves correspond to the values
of the parameter $q$. As seen, for small values of the planar angle deficit
the dependence of the FC on the magnetic flux is weak. In the right panel of
figure \ref{fig1} we have plotted the FC versus the temperature (in units of
$1/a$) for $r/a=1.5$. The numbers near the curves are the values of the
parameter $q$ and the full (dashed) curves correspond to $\alpha _{0}=0$ ($%
\alpha _{0}=0.4$). For $q=1$ (the curve between the full and dashed curves
for $q=3$) the dependence of the FC on $\alpha _{0}$ is weak and for that
case the full and dashed curves are almost the same. As seen from the
graphs, the dependence on the magnetic flux becomes weaker with decreasing
planar angle deficit (decreasing $q$). In accordance with the asymptotic
analysis given above, the suppression of the FC at high temperatures is seen
in the right panel.

\begin{figure}[tbph]
\begin{center}
\begin{tabular}{cc}
\epsfig{figure=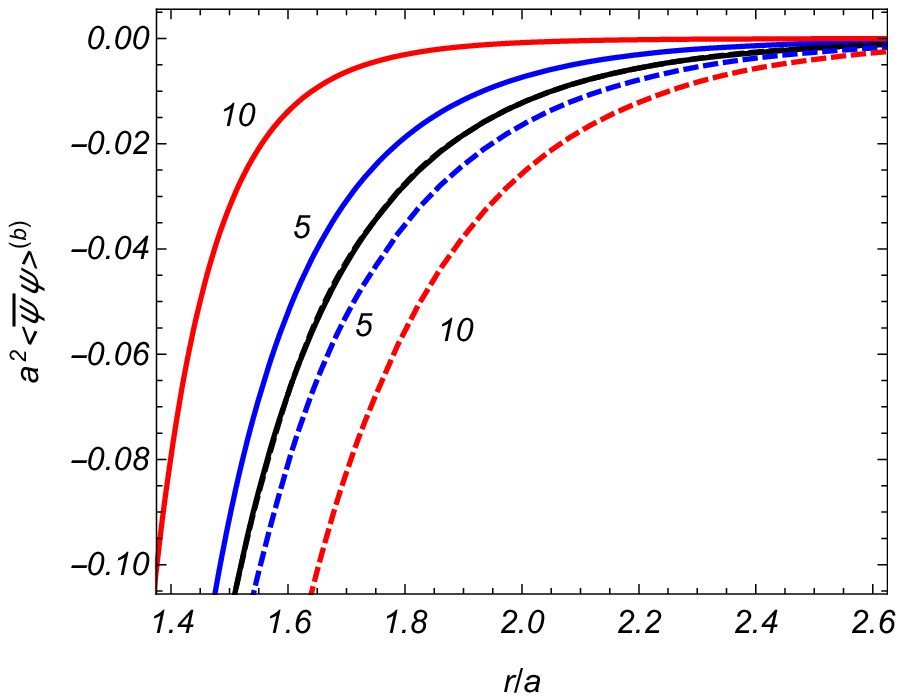,width=7.5cm,height=6.cm} & \quad %
\epsfig{figure=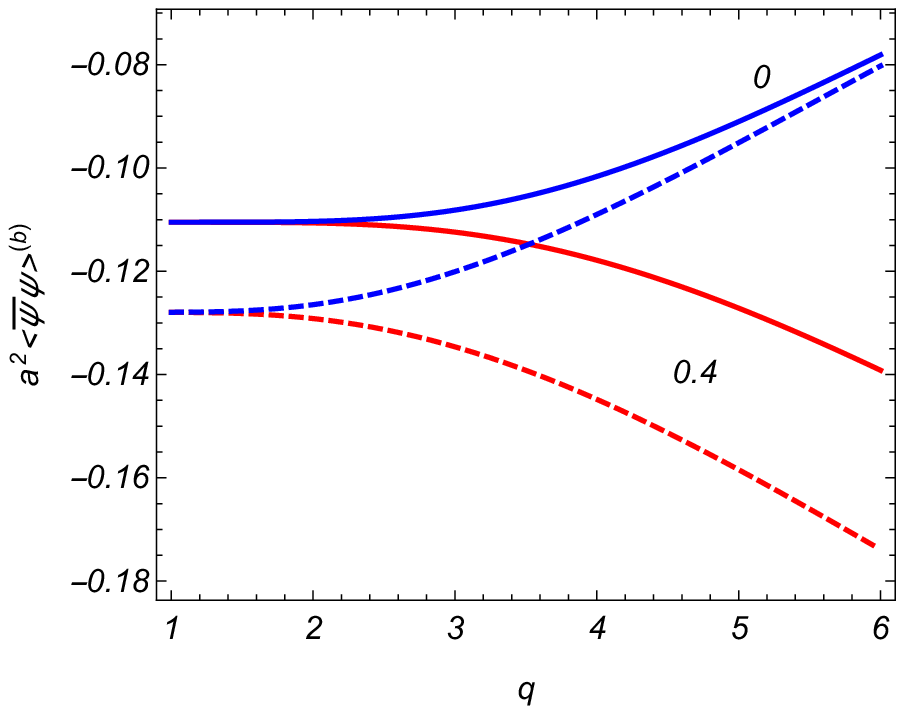,width=7.5cm,height=6.cm}%
\end{tabular}%
\end{center}
\caption{FC in the exterior region for a massless field with a zero chemical
potential as a function of the radial coordinate (left panel) and of the
parameter $q$ (right panel). For the left panel $Ta=0.5$ and the full
/dashed curves correspond to $\protect\alpha _{0}=0$/$\protect\alpha %
_{0}=0.4 $. The numbers near the curves are the values of $q$. The right
panel is plotted for $r/a=1.5$, $Ta=0.5$ (full curves), $Ta=0.25$ (dashed
curves) and the numbers near the curves are the corresponding values of $%
\protect\alpha _{0}$.}
\label{fig2}
\end{figure}

The dependence of the FC on the radial coordinate is shown in the left panel
of figure \ref{fig2} for fixed temperature corresponding to $Ta=0.5$. The
numbers near the curves present the values of the parameter $q$. The full
and dashed curves correspond to $\alpha _{0}=0$ and $\alpha _{0}=0.4$,
respectively. Again, for $q=1$ (the curves between the full and dashed
curves for $q=3$) the curves for $\alpha _{0}=0$ and $\alpha _{0}=0.4$ are
almost the same. As it has been shown above by the asymptotic analysis, for
large values of $Tr$ the FC is suppressed by the factor $e^{-2\pi Tr}$. The
dependence of the FC in the exterior region on the planar angle deficit is
displayed in the right panel of figure \ref{fig2} for $r/a=1.5$, $Ta=0.5$
(full curves) and $Ta=0.25$ (dashed curves). The figures near the curves
correspond to the values of the parameter $\alpha _{0}$. As seen from the
right panel, the behavior of the boundary-induced FC as a function of $q$ is
essentially different for $\alpha _{0}=0$ and $\alpha _{0}=0.4$.

Note that the boundary $r=a$ separates the exterior region from the region
where the magnetic flux is located. As a consequence of that, the results
presented in this section are valid for an arbitrary distribution of the
magnetic flux in the region $r<a$.

\section{FC inside a circular boundary}

\label{sec:Int}

In this section we consider the region $r\leqslant a$. The corresponding FC
in the vacuum state is presented as (\ref{FCvac2}), where the
boundary-induced contribution is given by \cite{Bell11}
\begin{eqnarray}
\langle \bar{\psi}\psi \rangle _{\mathrm{vac}}^{(b)} &=&-\frac{1}{\pi \phi
_{0}}\sum_{j}\int_{m}^{\infty }dx\,x\left\{ \mathrm{Im}\left[ \frac{\tilde{K}%
_{\beta _{j}}(xa)}{\tilde{I}_{\beta _{j}}(xa)}\right] \left[ I_{\beta
_{j}}^{2}(xr)+I_{\beta _{j}+\epsilon _{j}}^{2}(zr)\right] \right.  \notag \\
&&\left. +sm\mathrm{Re}\left[ \frac{\tilde{K}_{\beta _{j}}(xa)}{\tilde{I}%
_{\beta _{j}}(xa)}\right] \frac{I_{\beta _{j}+\epsilon
_{j}}^{2}(zr)-I_{\beta _{j}}^{2}(xr)}{\sqrt{x^{2}-m^{2}}}\right\} ,
\label{FCbvaci}
\end{eqnarray}%
with the notation for the modified Bessel functions
\begin{equation}
\tilde{F}_{\beta _{j}}(u)=uF_{\beta _{j}}^{\prime }(u)+\left( i\sqrt{%
u^{2}-m_{a}^{2}}+sm_{a}-\epsilon _{j}\beta _{j}\right) F_{\beta _{j}}(u),
\label{Ftild2}
\end{equation}%
where $F=I,K$.

Substituting the fermionic modes (\ref{psii}) in the mode-sum formula (\ref%
{FCpm}), for the contributions of the positive and negative energy modes to
the thermal part in the FC one gets%
\begin{equation}
\langle \bar{\psi}\psi \rangle _{T\lambda }=\lambda \frac{1}{2\phi _{0}a^{2}}%
\sum_{j}\sum_{l=1}^{\infty }\frac{T_{\beta _{j}}(\gamma _{j,l}^{(\lambda
)})g(\gamma _{j,l}^{(\lambda )})}{e^{\beta (E_{j,l}^{(\lambda )}-\lambda \mu
)}+1},  \label{FCTi}
\end{equation}%
where $\lambda =+,-$, $E_{j,l}^{(\lambda )}=\sqrt{\gamma _{j,l}^{(\lambda
)2}/a^{2}+m^{2}}$, $z=\gamma _{j,l}^{(\lambda )}$ are the positive zeros of
the function $\tilde{J}_{\beta _{j}}^{(\lambda )}(z)$ defined in accordance
with (\ref{Ftilde}) and we have introduced the notation%
\begin{equation}
g(z)=z\left[ \left( 1+\frac{\lambda sm_{a}}{\sqrt{z^{2}+m_{a}^{2}}}\right)
J_{\beta _{j}}^{2}(zr/a)-\left( 1-\frac{\lambda sm_{a}}{\sqrt{z^{2}+m_{a}^{2}%
}}\right) J_{\beta _{j}+\epsilon _{j}}^{2}(zr/a)\right] .  \label{gzi}
\end{equation}%
The roots $\gamma _{j,l}^{(\lambda )}$ are given implicitly and the
representation (\ref{FCTi}) is not convenient for the evaluation of the FC.
The summation formula for series of the type $\sum_{l=1}^{\infty }T_{\beta
_{j}}(\gamma _{j,l}^{(\lambda )})f(\gamma _{j,l}^{(\lambda )})$ has been
derived in \cite{Saha04} by using the generalized Abel-Plana formula from
\cite{Saharev,Saharev2} assuming that the function $f(z)$ is analytic in the
right half-plane of the complex variable $z$. In the problem at hand
\begin{equation}
f(z)=\frac{g(z)}{e^{\beta (\sqrt{z^{2}/a^{2}+m^{2}}-\lambda \mu )}+1},
\label{fz}
\end{equation}%
and for $\lambda \mu >0$ this function has simple poles $z=\gamma
_{n}^{(\lambda )}$, $n=0,\pm 1,\pm 2,\ldots $, in the right half-plane (the
case $\mu =0$ when the poles are located on the imaginary axis will be
discussed in appendix \ref{sec:appA}).

The procedure described in \cite{Saha04} can be generalized keeping the
terms in the generalized Abel-Plana formula coming from the poles in the
right half-plane. For the functions $f(x)$ real for real values of $x$, this
leads to the following summation formula
\begin{eqnarray}
\sum_{l=1}^{\infty }T_{\beta _{j}}(\gamma _{j,l}^{(\lambda )})f(\gamma
_{j,l}^{(\lambda )}) &=&\int_{0}^{\infty }dx\,f(x)+\frac{\pi }{2}\underset{%
z=0}{\mathrm{Res}}\frac{\tilde{Y}_{\beta _{j}}^{(\lambda )}(z)}{\tilde{J}%
_{\beta _{j}}^{(\lambda )}(z)}f(z)  \notag \\
&&-4\sum_{n=0}^{\infty }\mathrm{Re}\left[ e^{-i\pi \beta _{j}}\frac{\tilde{K}%
_{\beta _{j}}^{(\lambda )}(u_{n}^{(\lambda )})}{\tilde{I}_{\beta
_{j}}^{(\lambda )}(u_{n}^{(\lambda )})}\underset{z=iu_{n}^{(\lambda )}}{%
\mathrm{Res}}f(z)\right]  \notag \\
&&-\frac{2}{\pi }\int_{0}^{\infty }dx\,\mathrm{Re}\left[ e^{-\beta _{j}\pi
i}f(xe^{\pi i/2})\frac{\tilde{K}_{\beta _{j}}^{(\lambda )}(x)}{\tilde{I}%
_{\beta _{j}}^{(\lambda )}(x)}\right] ,  \label{SumAP}
\end{eqnarray}%
where $u_{n}^{(\lambda )}=-i\gamma _{n}^{(\lambda )}$ and for the modified
Bessel functions we have defined the notation%
\begin{equation}
\tilde{F}_{\beta _{j}}^{(\lambda )}(z)=zF_{\beta _{j}}^{\prime }(z)+\left(
\lambda \sqrt{\left( e^{\pi i/2}z\right) ^{2}+m_{a}^{2}}+sm_{a}-\epsilon
_{j}\beta _{j}\right) F_{\beta _{j}}(z),  \label{Ftild}
\end{equation}%
with $F=I,K$. The second term in the right-hand side of (\ref{SumAP}) comes
from the poles of the function $f(z)$ in the right half-plane. For an
analytic function $f(z)$ the formula (\ref{SumAP}) is reduced to the one in
\cite{Saha04}.

In the problem at hand the function $f(z)$ is given by (\ref{fz}). For this
function the integrand in the last term of (\ref{SumAP}) vanishes for $%
x<m_{a}$. The residue term at $z=0$ vanishes as well. The part in $\langle
\bar{\psi}\psi \rangle _{T\lambda }$ coming from the first integral on the
right-hand side of (\ref{SumAP}) presents the corresponding quantity in the
boundary-free geometry, denoted here as $\langle \bar{\psi}\psi \rangle
_{T\lambda }^{(0)}$. As a result, $\langle \bar{\psi}\psi \rangle _{T\lambda
}$ is presented as (\ref{FCpmb}), where for the boundary-induced
contribution (\ref{FCpmb}) one gets
\begin{eqnarray}
\langle \bar{\psi}\psi \rangle _{T\lambda }^{(b)} &=&\lambda \frac{1}{\pi
\phi _{0}}\sum_{j}\int_{m}^{\infty }dx\,\mathrm{Im}\left\{ \frac{\tilde{K}%
_{\beta _{j}}^{(\lambda )}(xa)}{\tilde{I}_{\beta _{j}}^{(\lambda )}(xa)}%
\frac{x}{e^{\beta (i\sqrt{x^{2}-m^{2}}-\lambda \mu )}+1}\right.  \notag \\
&&\times \left. \left[ \left( 1-\frac{\lambda ism}{\sqrt{x^{2}-m^{2}}}%
\right) I_{\beta _{j}}^{2}(xr)+\left( 1+\frac{\lambda ism}{\sqrt{x^{2}-m^{2}}%
}\right) I_{\beta _{j}+\epsilon _{j}}^{2}(zr)\right] \right\}  \notag \\
&&-\lambda \frac{2}{\phi _{0}}\theta \left( \lambda \mu \right)
T\sum_{j}\sum_{n=0}^{\infty }\mathrm{Im}\left\{ \frac{\tilde{K}_{\beta
_{j}}^{(\lambda )}(u_{n}^{(\lambda )}a)}{\tilde{I}_{\beta _{j}}^{(\lambda
)}(u_{n}^{(\lambda )}a)}\left[ (\pi \left( 2n+1\right) T-i\lambda \left( \mu
+sm\right) )I_{\beta _{j}}^{2}(u_{n}^{(\lambda )}r)\right. \right.  \notag \\
&&+\left.\left.(\pi \left( 2n+1\right) T-i\lambda \left( \mu -sm\right)
)I_{\beta_{j}+\epsilon _{j}}^{2}(u_{n}^{(\lambda )}r)\right]\right\}.
\label{FCb2i}
\end{eqnarray}%
The further transformation is similar to that for (\ref{FCb2}) with the
representation
\begin{eqnarray}
\langle \bar{\psi}\psi \rangle _{T\lambda }^{(b)} &=&\frac{1}{\pi \phi _{0}}%
\sum_{j}\int_{m}^{\infty }dx\,\mathrm{Im}\left\{ \frac{\tilde{K}_{\beta
_{j}}(xa)}{\tilde{I}_{\beta _{j}}(xa)}\frac{x}{e^{\lambda \beta (i\sqrt{%
x^{2}-m^{2}}-\mu )}+1}\right.  \notag \\
&&\times \left. \left[ \left( 1-\frac{ism}{\sqrt{x^{2}-m^{2}}}\right)
I_{\beta _{j}}^{2}(xr)+\left( 1+\frac{ism}{\sqrt{x^{2}-m^{2}}}\right)
I_{\beta _{j}+\epsilon _{j}}^{2}(zr)\right] \right\}  \notag \\
&&-\frac{2}{\phi _{0}}\theta \left( \lambda \mu \right)
T\sum_{j}\sum_{n=0}^{\infty }\mathrm{Im}\left\{ \frac{\tilde{K}_{\beta
_{j}}(u_{n}a)}{\tilde{I}_{\beta _{j}}(u_{n}a)}\left[ (\pi \left( 2n+1\right)
T-i\left( \mu +sm\right) )I_{\beta _{j}}^{2}(u_{n}r)\right. \right.  \notag
\\
&&\left. \left. +(\pi \left( 2n+1\right) T-i\left( \mu -sm\right) )I_{\beta
_{j}+\epsilon _{j}}^{2}(u_{n}r)\right] \right\} ,  \label{FCb3i}
\end{eqnarray}
where the notation with tilde is defined in accordance with (\ref{Ftild2}).

Summing the contributions from $\lambda =+$ and $\lambda =-$, we can see
that the sum of the first terms in the right-hand side of (\ref{FCb3i})
gives $-\langle \bar{\psi}\psi \rangle _{\mathrm{vac}}^{(b)}$. As a
consequence, for the boundary-induced contribution (\ref{FCbe}) one finds
\begin{eqnarray}
\langle \bar{\psi}\psi \rangle ^{(b)} &=&-\frac{2T}{\phi _{0}}%
\sum_{j}\sum_{n=0}^{\infty }\mathrm{Im}\left\{ \frac{\tilde{K}_{\beta
_{j}}(u_{n}a)}{\tilde{I}_{\beta _{j}}(u_{n}a)}\left[ (\pi \left( 2n+1\right)
T-i\left( \mu +sm\right) )I_{\beta _{j}}^{2}(u_{n}r)\right. \right.  \notag
\\
&&\left. \left. +(\pi \left( 2n+1\right) T-i\left( \mu -sm\right) )I_{\beta
_{j}+\epsilon _{j}}^{2}(u_{n}r)\right] \right\} ,  \label{FCbTi}
\end{eqnarray}%
where $u_{n}$ is defined by (\ref{un}). The total FC is presented as (\ref%
{FCdec2}). An equivalent representation for the notation in (\ref{FCbTi})\
is given by%
\begin{equation}
\tilde{F}_{\beta _{j}}(u)=\delta _{F}uF_{\beta _{j}+\epsilon _{j}}(u)+\left(
i\sqrt{u^{2}-m_{a}^{2}}+sm_{a}\right) F_{\beta _{j}}(u).  \label{Ftilden}
\end{equation}%
Similar to the case of the exterior region, we can see that $\langle \bar{%
\psi}\psi \rangle ^{(b)}$ is an even function under the simultaneous
reflections $\alpha \rightarrow -\alpha $, $\mu \rightarrow -\mu $. In (\ref%
{FCbTi}), the ratio of the modified Bessel functions can be presented in the
form%
\begin{equation}
\frac{\tilde{K}_{\beta _{j}}(z)}{\tilde{I}_{\beta _{j}}(z)}=\frac{W_{\beta
_{j},\beta _{j}+\epsilon _{j}}^{(+)}(z)+\left[ i\pi \left( 2n+1\right) T+\mu %
\right] a/z}{z[I_{\beta _{j}}^{2}(z)+I_{\beta _{j}+\epsilon
_{j}}^{2}(z)]+2sm_{a}I_{\beta _{j}}(z)I_{\beta _{j}+\epsilon _{j}}(z)},
\label{KIrat}
\end{equation}%
where $z=u_{n}a$ and $W_{\beta _{j},\beta _{j}+\epsilon _{j}}^{(+)}(z)$ is
defined by (\ref{Wpm}).

The FC in the case $\mu =0$ is considered in appendix \ref{sec:appA}. Though
the corresponding procedure for the evaluation of (\ref{FCbTi}) differs from
that we have described above for $\mu \neq 0$, the final result can be
obtained from (\ref{FCbTi}) taking the limit $\mu \rightarrow 0$:%
\begin{eqnarray}
\langle \bar{\psi}\psi \rangle ^{(b)} &=&-\frac{2T}{\phi _{0}}%
\sum_{j}\sum_{n=0}^{\infty }\left\{ \pi \left( 2n+1\right) T\mathrm{Im}\left[
\frac{\tilde{K}_{\beta _{j}}(u_{0n}a)}{\tilde{I}_{\beta _{j}}(u_{0n}a)}%
\right] \left[ I_{\beta _{j}}^{2}(u_{0n}r)+I_{\beta _{j}+\epsilon
_{j}}^{2}(u_{n}r)\right] \right.  \notag \\
&&\left. -sm\mathrm{Re}\left[ \frac{\tilde{K}_{\beta _{j}}(u_{0n}a)}{\tilde{I%
}_{\beta _{j}}(u_{0n}a)}\right] \left[ I_{\beta _{j}}^{2}(u_{0n}r)-I_{\beta
_{j}+\epsilon _{j}}^{2}(u_{0n}r)\right] \right\} ,  \label{FCbtoti0}
\end{eqnarray}%
with $u_{0n}$ from (\ref{u0n}). Note that now the arguments $u_{0n}a$ are
real and the imaginary and real parts in (\ref{FCbtoti0}) are directly
obtained from (\ref{KIrat}). For a massless field the expression for the
boundary-induced contribution in FC is reduced to%
\begin{equation}
\langle \bar{\psi}\psi \rangle ^{(b)}=-\frac{2T}{\phi _{0}a}%
\sum_{j}\sum_{n=0}^{\infty }\left. \frac{I_{\beta _{j}}^{2}(ur)+I_{\beta
_{j}+\epsilon _{j}}^{2}(ur)}{I_{\beta _{j}}^{2}(ua)+I_{\beta _{j}+\epsilon
_{j}}^{2}(ua)}\right\vert _{u=\pi \left( 2n+1\right) T},  \label{FCbm0i}
\end{equation}%
and it is negative. In this special case the boundary-free FC is zero and
the total FC is negative as well. For $2|\alpha _{0}|\leqslant 1-1/q$ the FC
given by (\ref{FCbm0i}) is a monotonically decreasing function of the radial
coordinate. That is not the case for $2|\alpha _{0}|>1-1/q$ when one of the
orders of the Bessel modified functions can be negative.

Now we return to a general case of the chemical potential and the mass and
consider the behavior of the boundary-induced FC (\ref{FCbTi}) near the cone
apex corresponding to small values of $r$. Redefining the summation variable
$j+n_{0}\rightarrow j$, the order $\beta _{j}$ of the modified Bessel
function is expressed in terms of $\alpha _{0}$. It can be seen that in the
limit $r\rightarrow 0$ the dominant contribution to the FC (\ref{FCbTi})
comes from the mode with $j=-\mathrm{sgn}(\alpha _{0})/2$. Expanding the
Bessel modified function for small values of the argument, to the leading
order we get
\begin{equation}
\langle \bar{\psi}\psi \rangle ^{(b)}\approx -\frac{qT(r/2)^{q-2q|\alpha
_{0}|-1}}{\pi \Gamma ^{2}((q+1)/2-q|\alpha _{0}|)}\sum_{n=0}^{\infty }%
\mathrm{Im}\left\{ \left[ \pi \left( 2n+1\right) T-i\left( \mu -\mathrm{sgn}%
(\alpha _{0})sm\right) \right] u_{n}^{2\beta _{j}}\frac{\tilde{K}_{\beta
_{j}}(u_{n}a)}{\tilde{I}_{\beta _{j}}(u_{n}a)}\right\} ,  \label{FCapex}
\end{equation}%
where $\beta _{j}=q\left( 1/2-|\alpha _{0}|\right) +\mathrm{sgn}(\alpha
_{0})/2$. As seen, the boundary-induced FC vanishes on the cone apex for $%
2|\alpha _{0}|<1-1/q$ and diverges for $2|\alpha _{0}|>1-1/q$. The
divergence in the latter case is related to the contribution of the
irregular mode at the cone apex. Note that, near the apex, for a massive
field the FC in the boundary-free geometry is dominated by the vacuum part
and the latter behaves as $1/r$ \cite{Bell16T}. Similar to the case of the
exterior region, it can be seen that for points near the boundary, under the
assumption $T(r-a)\ll 1$, the leading term in the asymptotic expansion over
the distance from the boundary coincides with that for the vacuum FC and
does not depend on the planar angle deficit and on the magnetic flux. It
diverges like $1/(a-r)^{2}$.

In the expressions for the FC in the exterior and interior regions, for the
modified Bessel functions $F_{\beta _{j}}(u)=I_{\beta _{j}}(u),K_{\beta
_{j}}(u)$, we have introduced the notations $\bar{F}_{\beta _{j}}(u)$ and $%
\tilde{F}_{\beta _{j}}(u)$. These notations are combined in a single
expression
\begin{eqnarray}
F_{\beta _{j}}^{(\eta )}(u) &=&uF_{\beta _{j}}^{\prime }(u)+\left[ \eta (i%
\sqrt{u^{2}-m_{a}^{2}}+sm_{a})-\epsilon _{j}\beta _{j}\right] F_{\beta
_{j}}(u)  \notag \\
&=&\delta _{F}uF_{\beta _{j}+\epsilon _{j}}(u)+\eta \left( i\sqrt{%
u^{2}-m_{a}^{2}}+sm_{a}\right) F_{\beta _{j}}(u),  \label{Feta}
\end{eqnarray}%
with $\eta =\pm 1$. For the normal to the boundary one $n_{\mu }=\eta \delta
_{\mu }^{1}$, where $\eta =+1$ in the interior region and $\eta =-1$ in the
exterior region, and $F_{\beta _{j}}^{(+1)}(u)=\tilde{F}_{\beta _{j}}(u)$, $%
F_{\beta _{j}}^{(-1)}(u)=\bar{F}_{\beta _{j}}(u)$.

\begin{figure}[tbph]
\begin{center}
\begin{tabular}{cc}
\epsfig{figure=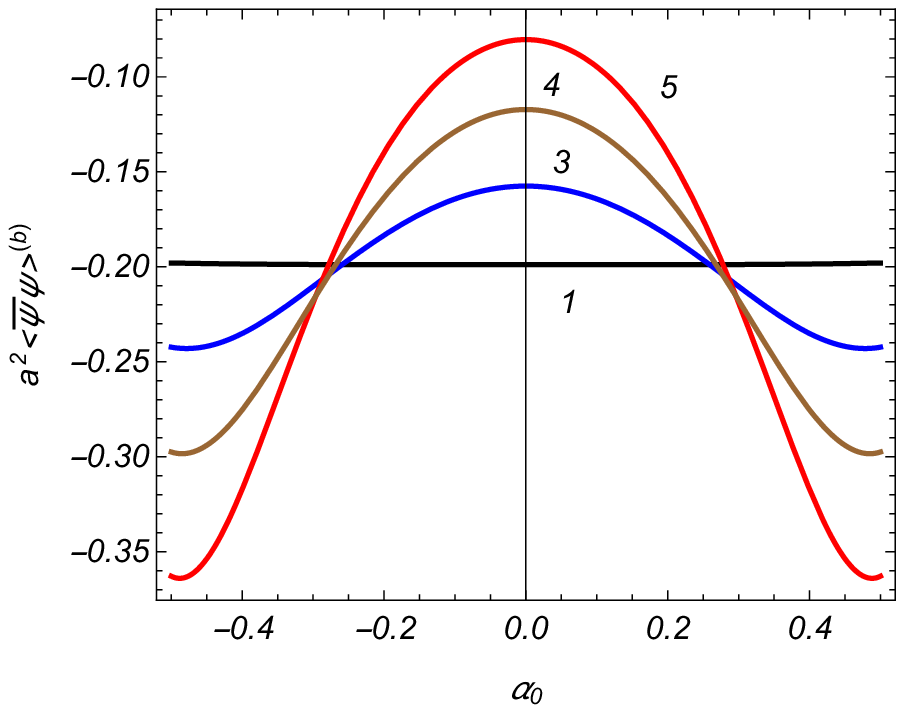,width=7.5cm,height=6.cm} & \quad %
\epsfig{figure=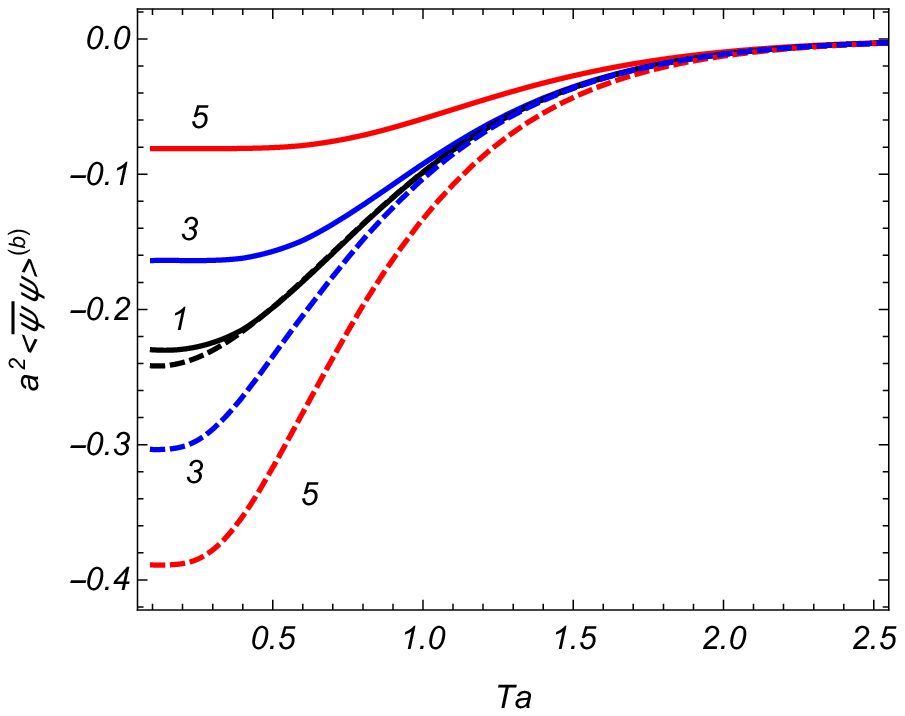,width=7.5cm,height=6.cm}%
\end{tabular}%
\end{center}
\caption{The same as in figure \protect\ref{fig1} for the interior region
with fixed $r/a=0.5$.}
\label{fig3}
\end{figure}

In figure \ref{fig3}, for a massless field with $\mu =0$, we have presented
the dependence of the FC inside a circular boundary on the parameter $\alpha
_{0}$ and on the temperature for fixed $r/a=0.5$. For the left panel $Ta=0.5$
and in the right panel the full and dashed curves correspond to $\alpha
_{0}=0$ and $\alpha _{0}=0.4$, respectively. On both the panels, the numbers
near the curves correspond to the values of $q$. Again, we see that for a
planar geometry, $q=1$, the dependence of the FC on the magnetic flux is
weak.

\begin{figure}[tbph]
\begin{center}
\begin{tabular}{cc}
\epsfig{figure=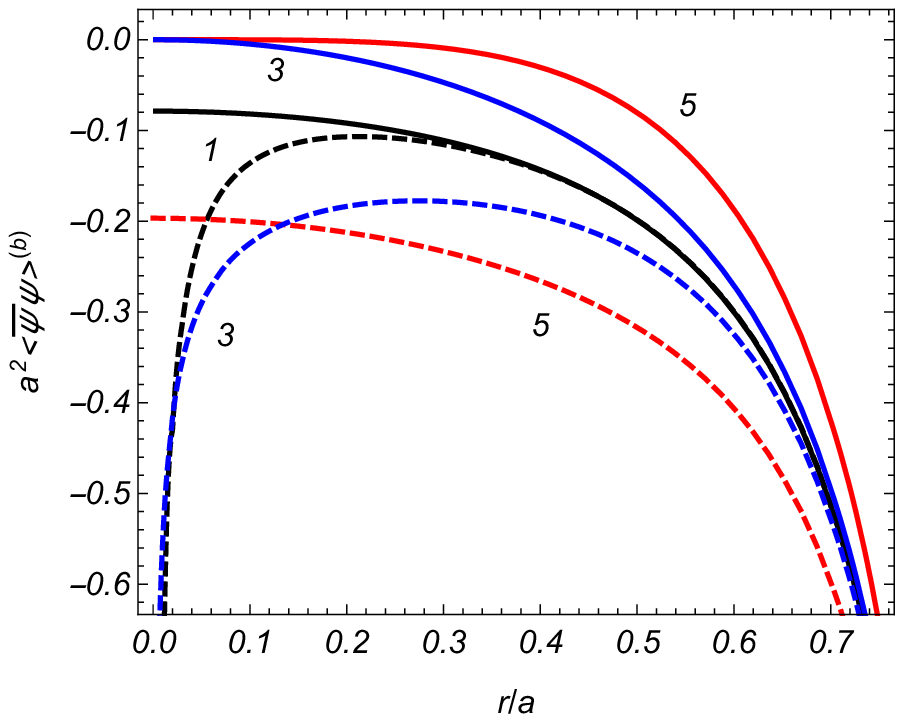,width=7.5cm,height=6.cm} & \quad %
\epsfig{figure=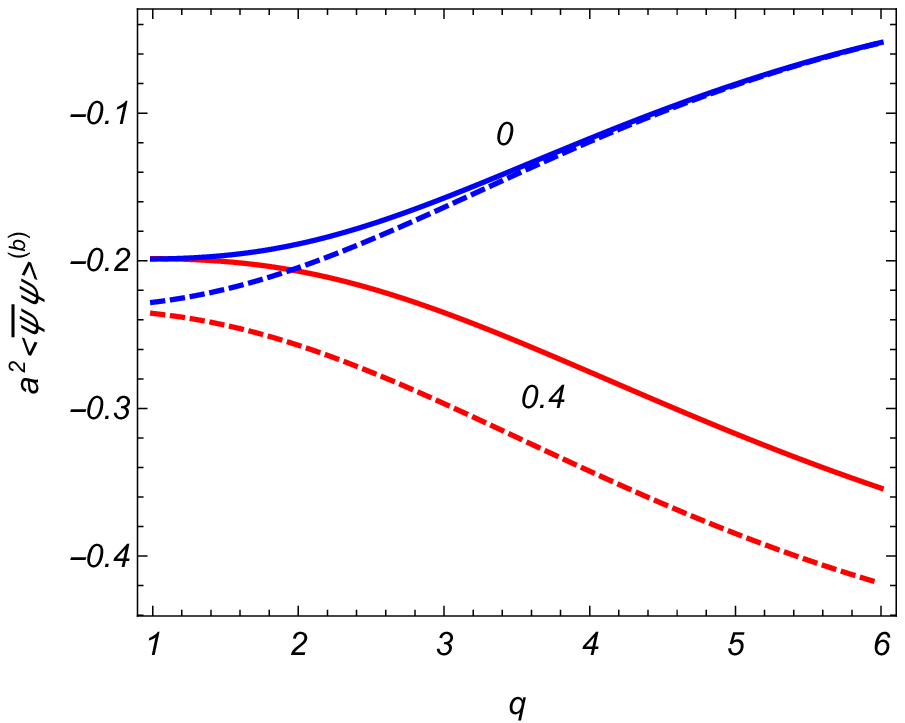,width=7.5cm,height=6.cm}%
\end{tabular}%
\end{center}
\caption{The same as in figure \protect\ref{fig2} for the interior region.
For the right panel we have taken $r/a=0.5$.}
\label{fig4}
\end{figure}

In figure \ref{fig4} we display the FC inside a circular boundary as a
function of the radial coordinate and of the parameter $q$. In the left
panel the numbers near the curves are the values of the parameter $q$, the
full/dashed curves correspond to $\alpha _{0}=0/\alpha _{0}=0.4$, and the
graphs are plotted for $Ta=0.5$. In the right panel $r/a=0.5$, $Ta=0.5$ for
full curves and $Ta=0.25$ for dashed curves. The numbers near the curves
correspond to the values of $\alpha _{0}$. As it has been already mentioned
above, in the case $2|\alpha _{0}|>1-1/q$ the FC in the interior region is
not a monotonic function of the radial coordinate.

\begin{figure}[tbph]
\begin{center}
\begin{tabular}{cc}
\epsfig{figure=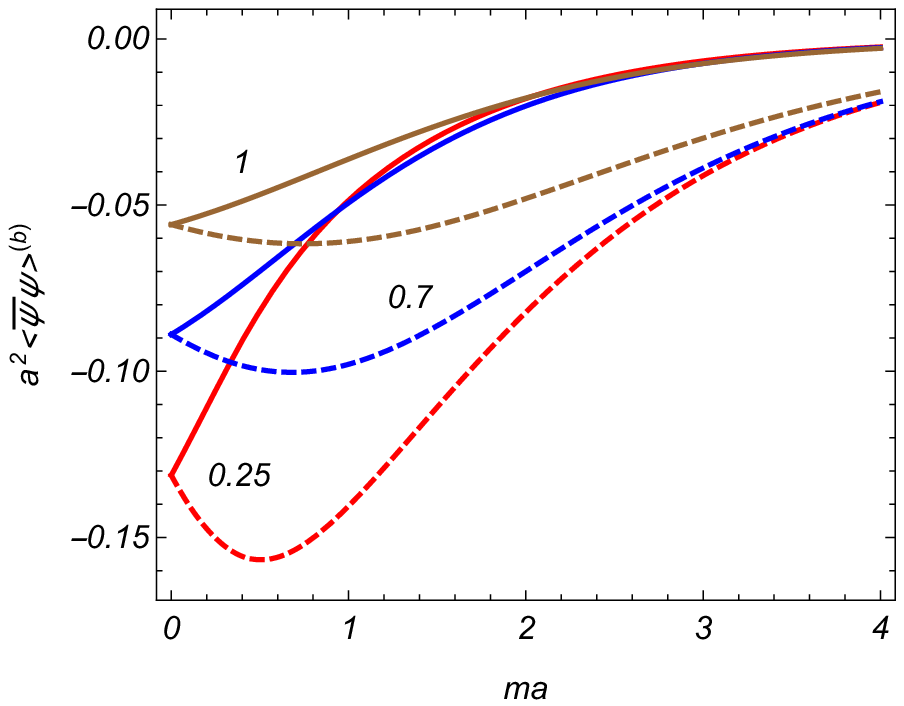,width=7.5cm,height=6.cm} & \quad %
\epsfig{figure=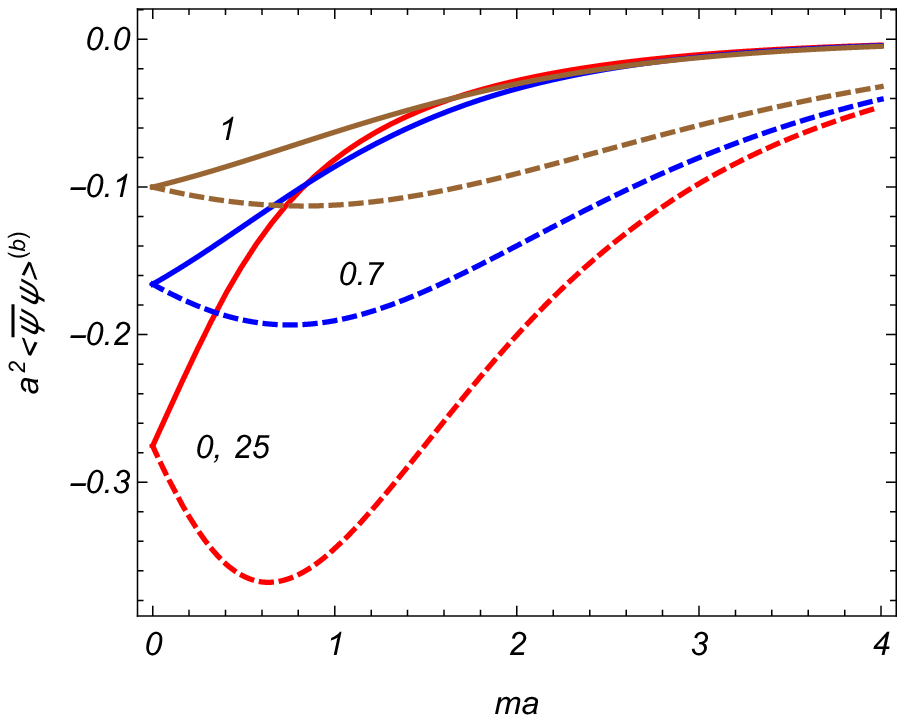,width=7.5cm,height=6.cm}%
\end{tabular}%
\end{center}
\caption{Boundary-induced FC as a function of the field mass in the case of
a zero chemical potential in the exterior (left panel, $r/a=1.5$) and
interior (right panel, $r/a=0.5$) regions. The full and dashed curves
correspond to $s=1$ and $s=-1$, respectively. The numbers near the curves
are the values of $Ta$ and for remaining parameters we have taken $q=2.5$, $%
\protect\alpha _{0}=0.4$.}
\label{fig5}
\end{figure}

In the numerical examples above we have considered the case of a massless
field. It is of interest to consider the effect of the mass on the FC. For a
massive field the FC will depend on the parameter $s$. For a field with zero
chemical potential, in figure \ref{fig5} we have plotted the dependence of
the FC on the mass outside (left panel) and inside (right panel) a circular
boundary for a conical space with $q=2.5$ and for the magnetic flux
corresponding to $\alpha _{0}=0.4$. The full and dashed curves correspond to
$s=1$ and $s=-1$, respectively, and the numbers near the curves are the
values of $Ta$. For the left and right panels we have taken $r/a=1.5$ and $%
r/a=0.5$, respectively. As is seen, the influence of the mass on the FC is
different for the cases $s=1$ and $s=-1$. In the first case the absolute
value of the FC decreases with increasing mass, whereas in the second case
the absolute value of the FC takes its maximum for some intermediate value
of the mass parameter. Note that figure \ref{fig5} presents the
boundary-induced contribution. For a massive field there is also nonzero
boundary-free part discussed in \cite{Bell16T}.

\begin{figure}[tbph]
\begin{center}
\begin{tabular}{cc}
\epsfig{figure=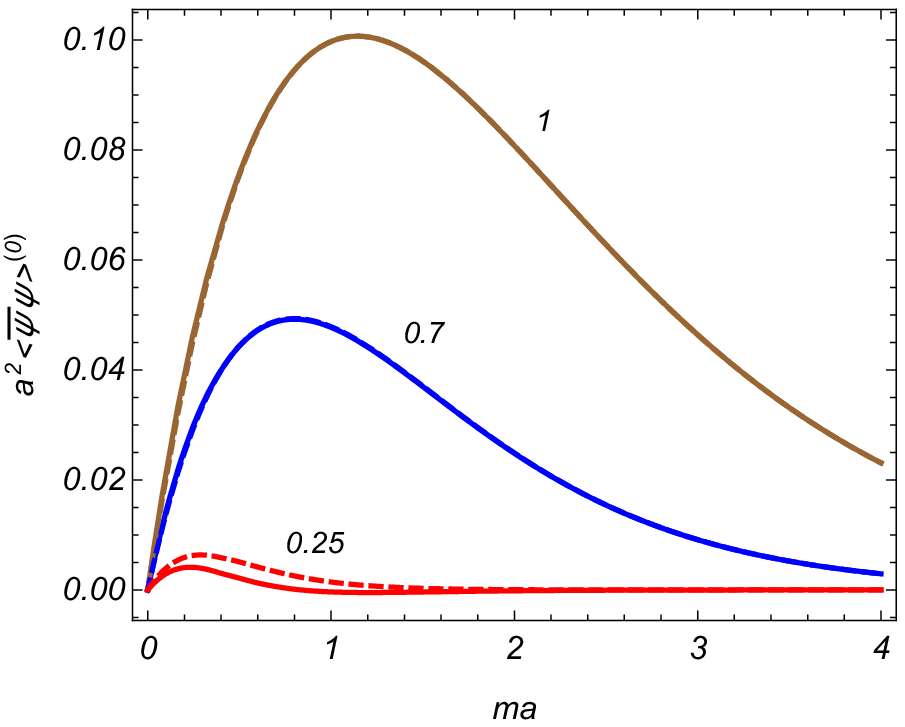,width=7.5cm,height=6.cm} & \quad %
\epsfig{figure=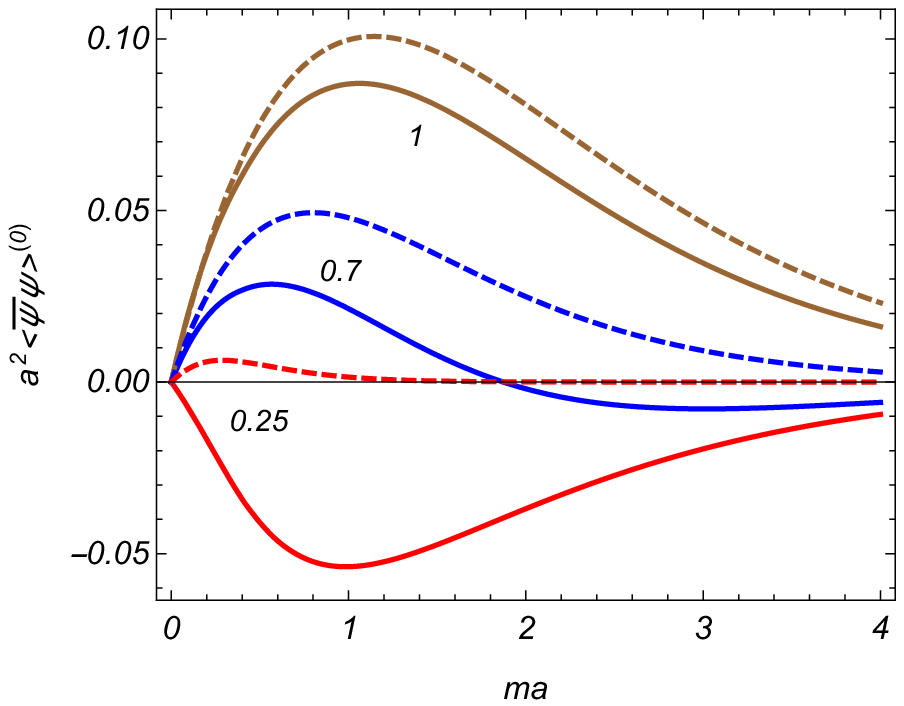,width=7.5cm,height=6.cm}%
\end{tabular}%
\end{center}
\caption{Boundary-free part in the FC (full curves) versus the field mass in
the case of a zero chemical potential and for the field with $s=1$. The left
and right panels are plotted for $r/a=1.5$ and $r/a=0.5$, respectively. The
dashed curves present the FC\ in (2+1)-dimensional Minkowski spacetime when
the magnetic flux is absent. The values of the remaining parameters are the
same as those for figure \protect\ref{fig5}.}
\label{fig6}
\end{figure}

In order to see the importance of the effects of a boundary on the finite
temperature FC, let us compare the boundary-induced FC, depicted in figure %
\ref{fig5} with the corresponding quantity in the boundary-free conical
space. First of all, we note that for a massless field the FC\ in the
boundary-free geometry vanishes and the nonzero FC is a purely
boundary-induced effect. In this case, the influence of the finite
temperature on the FC is seen from figures \ref{fig1} and \ref{fig3}. For a
massive field the boundary-free part of the FC is given by Eq. (\ref{FC0n})
with separate contributions from Eqs. (\ref{FC0M}) and (\ref{FC03}). This
part has opposite signs for the cases $s=1$ and $s=-1$. In figure \ref{fig6}
we have displayed the boundary-free FC (full curves) for the case $s=1$ as a
function of the field mass. The left and right panels correspond to $r/a=1.5$
and $r/a=0.5$ and in the numerical evaluation we have taken $q=2.5$ and $%
\alpha _{0}=0.4$ (the same as in figure \ref{fig5}). Similar to figure \ref%
{fig5}, the numbers near the curves correspond to the values of $Ta$. The
dashed curves correspond to the quantity $a^{2}\langle \bar{\psi}\psi
\rangle _{\mathrm{M}}^{(0)}$, where the FC $\langle \bar{\psi}\psi \rangle _{%
\mathrm{M}}^{(0)}$ in (2+1)-dimensional Minkowski spacetime, in the absence
of the magnetic flux, is given by (\ref{FC0M}) with $s=1$. The FC $\langle
\bar{\psi}\psi \rangle _{\mathrm{M}}^{(0)}$ does not depend on the radial
coordinate and the dashed curves on the left and right panels coincide. At
large distances from the cone apex the relative contribution of the
topological part is small, whereas near the apex it is essential. As seen
from the graphs, the topological contribution may change the sign of the FC
(the graph for $Ta=0.25$ on the left panel and the graphs for $Ta=0.25,0.7$
on the right panel). Now, comparing the graphs in figures \ref{fig5} and \ref%
{fig6}, we see that, for the values of the parameters used in the numerical
evaluation, the boundary-induced contributions to the finite temperature FC\
are essential and they may qualitatively change the behavior of the FC.

The geometry inside a circular boundary, discussed in this section, can be
considered as a limiting case of a conical ring with a fermionic field
localized in the region $b<r<a$ and obeying the MIT bag boundary condition (%
\ref{BCMIT}) on the edges $r=a,b$. Similar to the limiting transition $%
a\rightarrow 0$, discussed in the previous section, we expect that for fixed
$r$ and $|\alpha _{0}|\neq 1/2$, the contribution of the boundary at $r=b$
to the FC\ will tend to zero in the limit $b\rightarrow 0$. Consequently,
for $Tb,mb,b/r\ll 1$, the results of this section will approximate the FC in
conical rings threaded by a magnetic flux.

We could consider the boundary condition
\begin{equation}
\left( 1-in_{\mu }\gamma ^{\mu }\right) \psi (x)=0,  \label{BCMIT2}
\end{equation}%
that differs from (\ref{BCMIT}) by the sign of the term containing the
normal to the boundary. As it has been already noticed in \cite{Berr87},
this type of condition is an equally acceptable for the Dirac equation. The
mode functions for the case of boundary condition (\ref{BCMIT2}) are
obtained from the mode functions (\ref{psie}) and (\ref{psii}) by changing
the signs of the terms with $\sqrt{z^{2}+m_{a}^{2}}$ and $sm_{a}$ in the
definitions of the notations (\ref{Fbar}) and (\ref{Ftilde}). The final
formulas for the boundary-induced contribution in the FC, $\langle \bar{\psi}%
\psi \rangle ^{(b)}$, are obtained from (\ref{FCb4}) and (\ref{FCbTi}) by
changing the signs of the terms with $\sqrt{u^{2}-m_{a}^{2}}$ and $sm_{a}$
in the notations (\ref{FbarIK}) and (\ref{Ftild2}). Note that this
corresponds to the change $\eta \rightarrow -\eta $ in (\ref{Feta}). Let us
denote the boundary-induced FC for given $s$ and $\mu $ in the cases of the
boundary conditions (\ref{BCMIT}) and (\ref{BCMIT2}) by $\langle \bar{\psi}%
\psi \rangle _{s}^{(b,+1)}(\mu )$ and $\langle \bar{\psi}\psi \rangle
_{s}^{(b,-1)}(\mu )$, respectively. The corresponding expressions can be
written in combined form%
\begin{eqnarray}
\langle \bar{\psi}\psi \rangle _{s}^{(b,\eta )}(\mu ) &=&-\frac{2T}{\phi _{0}%
}\sum_{j}\sum_{n=0}^{\infty }\mathrm{Im}\left\{ \frac{K_{\beta _{j}}^{(\eta
)}(u_{n}a)}{I_{\beta _{j}}^{(\eta )}(u_{n}a)}\left[ (\pi \left( 2n+1\right)
T-i\left( \mu +sm\right) )I_{\beta _{j}}^{2}(u_{n}r)\right. \right.  \notag
\\
&&\left. \left. +(\pi \left( 2n+1\right) T-i\left( \mu -sm\right) )I_{\beta
_{j}+\epsilon _{j}}^{2}(u_{n}r)\right] \right\} ,  \label{FCsi}
\end{eqnarray}%
in the interior region and
\begin{eqnarray}
\langle \bar{\psi}\psi \rangle _{s}^{(b,\eta )}(\mu ) &=&-\frac{2T}{\phi _{0}%
}\sum_{j}\sum_{n=0}^{\infty }\mathrm{Im}\left\{ \frac{I_{\beta _{j}}^{(-\eta
)}(u_{n}a)}{K_{\beta _{j}}^{(-\eta )}(u_{n}a)}\left[ (\pi \left( 2n+1\right)
T-i\left( \mu +sm\right) )K_{\beta _{j}}^{2}(u_{n}r)\right. \right.  \notag
\\
&&\left. \left. +(\pi \left( 2n+1\right) T-i\left( \mu -sm\right) )K_{\beta
_{j}+\epsilon _{j}}^{2}(u_{n}r)\right] \right\} .  \label{FCse}
\end{eqnarray}%
in the exterior region. Here $\eta $ specifies the boundary condition: $\eta
=+1$ for (\ref{BCMIT}) and $\eta =-1$ for (\ref{BCMIT2}). On the base of
these formulas, by taking into account the relations%
\begin{equation}
\left. F_{\beta _{j}}^{(-\eta )}(u_{n}(-\mu )a)\right\vert _{s=\pm 1}=\left[
F_{\beta _{j}}^{(\eta )}(u_{n}(\mu )a)\right] _{s=\mp 1}^{\ast },
\label{RelFeta}
\end{equation}%
we see that%
\begin{equation}
\langle \bar{\psi}\psi \rangle _{s}^{(b,-1)}(\mu )=-\langle \bar{\psi}\psi
\rangle _{-s}^{(b,+1)}(-\mu ).  \label{RelFC}
\end{equation}%
This gives the relation between the boundary-induced FCs for the boundary
conditions (\ref{BCMIT}) and (\ref{BCMIT2}). In particular, for a massless
field with zero chemical potential the FC is given by (\ref{FCbm0}) and (\ref%
{FCbm0i}) for the boundary condition (\ref{BCMIT}) and by the same
expressions with the opposite signs for the condition (\ref{BCMIT2}).

\section{FC in parity and time-reversal symmetric models}

\label{sec:PT}

As it has been mentioned above, in (2+1) dimensions one has two inequivalent
irreducible representations of the Clifford algebra. These representations
can be realized by two sets of $2\times 2$ gamma matrices $\gamma
_{(s)}^{\mu }=(\gamma ^{0},\gamma ^{1},\gamma _{(s)}^{2}=-is\gamma
^{0}\gamma ^{1}/r)$, where $\gamma ^{0}$ and $\gamma ^{1}$ are given by (\ref%
{gamma}) and $s=\pm 1$. The representation with $s=+1$ corresponds to $%
\gamma ^{2}$ in (\ref{gamma}). In separate representations with given $s$,
the mass term in the Lagrangian density $L_{s}=\bar{\psi}_{(s)}(i\gamma
_{(s)}^{\mu }D_{\mu }-m)\psi _{(s)}$ breaks the parity ($P$) and
time-reversal ($T$) invariances of the fermionic model. In the absence of
magnetic fields, $P$- and $T$-invariant models in (2+1) dimensions can be
constructed considering a set of two fields $\psi _{(+1)}$ and $\psi _{(-1)}$
with the Lagrangian density $L=\sum_{s=\pm 1}L_{s}$. First let us consider
the case when both the fields obey the boundary condition (\ref{BCMIT}) on
the circle $r=a$:
\begin{equation}
(1+in_{\mu }\gamma _{(s)}^{\mu })\psi _{(s)}(x)=0.  \label{BC1}
\end{equation}%
We can formulate the model in terms of new fields $\psi _{(s)}^{\prime }$
defined as $\psi _{(+1)}^{\prime }=\psi _{(+1)}$ and $\psi _{(-1)}^{\prime
}=\gamma ^{0}\gamma ^{1}\psi _{(-1)}$. The Lagrangian density is presented
as $L=\sum_{s=\pm 1}\bar{\psi}_{(s)}^{\prime }(i\gamma ^{\mu }D_{\mu
}-sm)\psi _{(s)}^{\prime }$ with the gamma matrices defined by (\ref{gamma})
and the Dirac equation for the separate fields is in the form (\ref{Dirac}).
The boundary conditions for new fields take the form $\left( 1+isn_{\mu
}\gamma ^{\mu }\right) \psi _{(s)}^{\prime }(x)=0$. Introducing 4-component
spinor $\Psi =(\psi _{(+1)}^{\prime },\psi _{(-1)}^{\prime })^{T}$ and $%
4\times 4$ Dirac matrices $\gamma _{(4)}^{\mu }=\sigma _{3}\otimes \gamma
^{\mu }$, with $\sigma _{3}=\mathrm{diag}(1,-1)$, the Lagrangian density is
written as $L=\bar{\Psi}(i\gamma _{(4)}^{\mu }D_{\mu }-m)\Psi $ with the
boundary condition $(1+in_{\mu }\gamma _{(4)}^{\mu })\Psi (x)=0$ on $r=a$.
The latter is the bag boundary condition for the 4-component spinor.

For the FC corresponding to the field $\psi _{(s)}$ one has $\langle \bar{%
\psi}_{(s)}\psi _{(s)}\rangle =s\langle \bar{\psi}_{(s)}^{\prime }\psi
_{(s)}^{\prime }\rangle $ and for the total FC we get
\begin{equation}
\langle \bar{\Psi}\Psi \rangle =\sum_{s=\pm 1}\langle \bar{\psi}_{(s)}\psi
_{(s)}\rangle =\sum_{s=\pm 1}s\langle \bar{\psi}_{(s)}^{\prime }\psi
_{(s)}^{\prime }\rangle .  \label{FCtot}
\end{equation}%
The expressions for the separate terms in the last sum of (\ref{FCtot}) are
obtained from the results of the previous sections. The field $\psi
_{(+1)}^{\prime }$ obeys the same equation and the boundary condition (the
condition (\ref{BCMIT})) as the field $\psi (x)$ in section \ref{sec:modes}
with $s=+1$ and the boundary-induced contribution to the corresponding FC in
the interior and exterior regions is given by (\ref{FCsi}) and (\ref{FCse})
with $s=1$ and $\eta =+1$. The field $\psi _{(-1)}^{\prime }$ obeys the same
equation as the field $\psi (x)$ with $s=-1$ and the boundary condition (\ref%
{BCMIT2}). The corresponding boundary-induced contribution to the FC\ is
given by (\ref{FCsi}) and (\ref{FCse}) with $s=-1$ and $\eta =-1$. By taking
into account the relation (\ref{RelFC}) we see that%
\begin{equation}
\langle \bar{\psi}_{(-1)}^{\prime }\psi _{(-1)}^{\prime }\rangle ^{(b)}(\mu
)=-\langle \bar{\psi}_{(+1)}^{\prime }\psi _{(+1)}^{\prime }\rangle
^{(b)}(-\mu ).  \label{RelFC1}
\end{equation}%
Hence, the boundary-induced contribution to the total FC is presented in the
form%
\begin{equation}
\langle \bar{\Psi}\Psi \rangle ^{(b)}=\sum_{l=\pm 1}\langle \bar{\psi}%
_{(+1)}^{\prime }\psi _{(+1)}^{\prime }\rangle ^{(b)}(l\mu ),  \label{FCtot2}
\end{equation}%
where $\langle \bar{\psi}_{(+1)}^{\prime }\psi _{(+1)}^{\prime }\rangle
^{(b)}(\mu )$ is given by (\ref{FCsi}) and (\ref{FCse}) with $s=1$ and $\eta
=+1$. By taking into account that $\langle \bar{\psi}_{(+1)}^{\prime }\psi
_{(+1)}^{\prime }\rangle ^{(b)}(\mu )$ is an even function under the
transformation $\alpha \rightarrow -\alpha $, $\mu \rightarrow -\mu $, from (%
\ref{FCtot2}) it follows that the FC $\langle \bar{\Psi}\Psi \rangle ^{(b)}$
is an even function of $\mu $ and $\alpha $ separately. In the case of the
zero chemical potential, $\mu =0$, we get $\langle \bar{\Psi}\Psi \rangle
^{(b)}=2\langle \bar{\psi}_{(+1)}^{\prime }\psi _{(+1)}^{\prime }\rangle
^{(b)}$.

We could consider the case when the fields $\psi _{(s)}$ with $s=+1$ and $%
s=-1$ obey different boundary conditions: $(1+isn_{\mu }\gamma _{(s)}^{\mu
})\psi _{(s)}(x)=0$, $r=a$. This type of problem has been discussed in \cite%
{Rech07} for graphene rings, where the parameter $s$ corresponds to valley
degree of freedom (see below). In this case, the transformed fields $\psi
_{(s)}^{\prime }(x)$ obey the same boundary condition $\left( 1+in_{\mu
}\gamma ^{\mu }\right) \psi _{(s)}^{\prime }(x)=0$. The corresponding
condensate $\langle \bar{\psi}_{(s)}^{\prime }\psi _{(s)}^{\prime }\rangle
^{(b)}$ is given by (\ref{FCsi}) and (\ref{FCse}) with $\eta =+1$ and the
total FC is obtained by using the last relation in (\ref{FCtot}). For a
massless field one has $\langle \bar{\psi}_{(-1)}^{\prime }\psi
_{(-1)}^{\prime }\rangle ^{(b)}=\langle \bar{\psi}_{(+1)}^{\prime }\psi
_{(+1)}^{\prime }\rangle ^{(b)}$ and the total FC vanishes $\langle \bar{\Psi%
}\Psi \rangle ^{(b)}=0$.

Among the condensed matter realizations of the fermionic model, we have
considered, are graphitic cones (carbon nanocones in another terminology).
The long-wavelength properties of the corresponding electronic subsystem are
well described by a set of two-component spinors $(\psi _{(+1)},\psi
_{(-1)}) $, obeying the Dirac equation with the speed of light replaced by
the Fermi velocity of electrons (see, for example, \cite{Gusy07}). These
spinors correspond to the two different inequivalent points $\mathbf{K}_{+}$
and $\mathbf{K}_{-}$ at the corners of the two-dimensional Brillouin zone
for the graphene hexagonal lattice. The parameter $s=\pm 1$ in the
discussion above corresponds to valley degree of freedom in graphene. The
components of the separate spinors $\psi _{(s)}$ give the amplitude of the
electron wave function on triangular sublattices $A$ and $B$ of the graphene
hexagonal lattice. Graphitic cones are obtained from planar graphene sheet
if one or more sectors with the angle $\pi /3$ are excised and the remainder
is joined. The opening angle of the cone is given by $\phi _{0}=2\pi
(1-n_{c}/6)$, where $n_{c}=1,2,\ldots ,5$ is the number of the removed
sectors. The graphitic cones with these values of opening angle have been
experimentally observed \cite{Kris97}. The electronic structure of graphitic
cones was investigated in \cite{Lamm00}-\cite{Site08}. Note that the
graphitic cones have been observed in both the forms as caps on the ends of
the nanotubes and as free-standing structures (see, for instance, \cite%
{Char01} and references therein). The geometry outside the circular
boundary, we have considered above, corresponds to the continuum description
of graphitic cones with cutted apex. As it has been discussed in \cite%
{Lamm00}, that can be done with acid or with an STM. For even values of $%
n_{c}$ the periodicity condition for 4-spinor $\Psi =(\psi _{(+1)},\psi
_{(-1)})^{T}$ under the rotation around the cone apex has the form $\Psi
(t,r,\phi +\phi _{0})=-\cos \left( \pi n_{c}/2\right) \Psi (t,r,\phi )$ and
it does not mix the spinors $\psi _{(+1)}$ and $\psi _{(-1)}$. For $n_{c}=2$
this corresponds to the condition we have discussed in preceding sections
and the corresponding FC is obtained by combining the contributions from $%
s=+1$ and $s=-1$ in the way we have described above. For $n_{c}=4$ one has
an antiperiodic boundary condition and for the parameter $\chi $ in (\ref%
{QPC}) we get $\chi =1/2$. In this case the FC for separate fields $\psi
_{(s)}$ are obtained from the formulae given in the preceding sections by
the replacement $\alpha \rightarrow \alpha +1/2$. By a gauge transformation
this can be interpreted as a shift in the magnetic flux. Note that the Dirac
mass $m$ in the formulae given above is expressed in terms of the energy gap
$\Delta $ in graphene by the relation $m=\Delta /v_{F}^{2}$, where $%
v_{F}\approx 7.9\times 10^{7}$cm/s is the Fermi velocity of electrons.
Depending on the gap generation mechanism, the energy gap varies in the
range $1\,\mathrm{meV}\lesssim \Delta \lesssim 1\,\mathrm{eV}$.

\section{Conclusion}

\label{sec:Conc}

We have considered the combined effects of finite temperature and circular
boundary on the FC in a (2+1)-dimensional conical spacetime with an
arbitrary value of the planar angle deficit. Two types of boundary
conditions were used. The first one corresponds to the MIT bag boundary
condition and the second one, given by (\ref{BCMIT2}), differs by the sign
in front of the term containing the normal to the boundary. In
two-dimensional spaces there exist two inequivalent representations of the
Clifford algebra and we have presented the investigation for both the fields
realizing those representations. For the evaluation of the FC, the direct
summation over a complete set of fermionic modes is employed. In the case of
the bag boundary condition those modes outside and inside the circular
boundary are given by (\ref{psie}) and (\ref{psii}). In the region inside
the circular boundary the eigenvalues of the radial quantum number $\gamma $
are roots of the equation (\ref{modesi}). They are given implicitly and for
the summation of the corresponding series in the mode sum we have
generalized the formula from \cite{Saha04} for functions having poles in the
right-half plane. That allowed us to present the FC in the form where the
explicit knowledge of the eigenvalues for $\gamma $ is not required.

The FCs in both the exterior and interior regions are decomposed into
boundary-free and boundary-induced contributions, as given by (\ref{FCdec2}%
). The boundary-free geometry has been discussed in \cite{Bell16T} and we
were mainly concerned with the effects induced by the boundary. For a
general case of a massive fermionic field with nonzero chemical potential,
the boundary-induced contributions in the exterior and interior regions are
given by expressions (\ref{FCb4}) and (\ref{FCbTi}). They are periodic
functions of the magnetic flux with the period equal to the flux quantum and
even functions under the simultaneous reflections $\alpha \rightarrow
-\alpha $, $\mu \rightarrow -\mu $. The expressions for the boundary-induced
FCs are further simplified for a field with zero chemical potential (see (%
\ref{FCbtotmu0}) and (\ref{FCbtoti0})). For a massless field they are
reduced to (\ref{FCbm0}) and (\ref{FCbm0i}). The dependence of the FC on the
magnetic flux becomes weaker with decreasing planar angle deficit. For
points near the boundary, the contribution of the high-energy modes dominate
in the expectation values and the leading term in the asymptotic expansion
over the distance from the boundary coincides with that for the vacuum FC.
In this region the effects of finite temperature, of planar angle deficit
and of magnetic flux are weak. As expected, at large distances from the
boundary the FC is dominated by the Minkowskian term $\langle \bar{\psi}\psi
\rangle _{\mathrm{M}}^{(0)}$, given by (\ref{FC0M}). For $Tr\gg 1$ the
boundary-induced FC is exponentially suppressed. Similar behavior takes
place for the topological part in the boundary-free FC. The behavior of the
boundary-induced FC near the cone apex critically depends on the magnetic
flux and on the planar angle deficit. It vanishes on the cone apex for $%
2|\alpha _{0}|<1-1/q$ and diverges for $2|\alpha _{0}|>1-1/q$. The
divergence is related to the presence of the mode irregular at the cone
apex. For a fixed distance from the boundary and at high temperatures the FC
is dominated by the Minkowskian part.

We have also considered the FC for the boundary condition (\ref{BCMIT2})
that differs from the condition (\ref{BCMIT}) by the sign of the term with
the normal to the boundary. The corresponding formulas are obtained from
those for the condition (\ref{BCMIT}) by using the relations (\ref{RelFC}).
In the special case of a massless field with zero chemical potential the FCs
for the boundary conditions (\ref{BCMIT}) and (\ref{BCMIT2}) differ by the
sign only.

For a fermionic field realizing a irreducible representation of the Clifford
algebra, the mass term breaks the $P$- and $T$- invariances. In order to
construct $P$- and $T$- invariant models one can combine two fields
corresponding to inequivalent representations. If both the fields obey the
boundary condition (\ref{BCMIT}), the boundary-induced contribution in the
total FC for this type of models is obtained from the results discussed in
sections \ref{sec:Ext} and \ref{sec:Int} by using the relation (\ref{FCtot2}%
) and it is an even function of the chemical potential and of the parameter $%
\alpha $. Another possibility corresponds to the situation when the fields
in different irreducible representations obey boundary conditions with
different signs of the term involving the normal to the boundary. In this
case the total FC is obtained with the help of the first relation in (\ref%
{FCtot2}) where the separate terms are directly taken from the results in
sections \ref{sec:Ext} and \ref{sec:Int} for the boundary condition (\ref%
{BCMIT}). For a massless field the parts $\langle \bar{\psi}_{(s)}^{\prime
}\psi _{(s)}^{\prime }\rangle ^{(b)}$ do not depend on the parameter $s$ and
the total FC is zero. From the results presented in the present paper the FC
can be obtained in graphitic cones with edges for the values of the opening
angle $\phi _{0}=2\pi (1-n_{c}/6)$ corresponding to even values of $n_{c}$
(the number of the sectors with the angle $\pi /3$, excised from planar
graphene).

\section*{Acknowledgments}

Aram Saharian was supported by CNPq Program APV. Process n. 453571/2018-2.
E. R. B. M. was partially supported by CNPq. through the project No.
313137/2014-5. Astghik Saharyan acknowledges additional support from the
European Union's Horizon 2020 research and innovation program under the
Marie Sklodowska-Curie grant agreement No. 765075 (LIMQUET).

\appendix

\section{Zero chemical potential}

\label{sec:appA}

In this appendix we consider the transformation of the mode-sum for the FC
in the case of the zero chemical potential, $\mu =0$. First we consider the
exterior region, $r>a$. The boundary-induced contribution in the thermal
part of the FC is given by the expression (\ref{FCpmb1}) with $\mu =0$. Now
the poles of the integrand are located on the imaginary axis:%
\begin{equation}
E=E_{n}\equiv i\pi \left( 2n+1\right) T,\;n=0,\pm 1,\pm 2,\ldots ,
\label{En0}
\end{equation}%
with $n=0,\pm 1,\pm 2,\ldots $. For the values of $\gamma =\gamma _{n}$
corresponding to the poles (\ref{En0}) in the upper half-plane one has
\begin{equation}
\gamma _{n}=iu_{0n},\;n=0,1,2,\ldots ,  \label{gamn0}
\end{equation}%
with $u_{0n}$ defined in (\ref{u0n}), and for the poles in the lower
half-plane we get $E_{n}^{(\lambda )}=E_{-n-1}^{(\lambda )\ast }$, $\gamma
_{n}=\gamma _{-n-1}^{\ast }$,$\;n=\ldots ,-2,-1$. As the next step, we
rotate the integration contour in (\ref{FCpmb1}) by the angle $\pi /2$ for $%
l=1$ and by the angle $-\pi /2$ for $l=2$. The poles $\gamma _{n}$, $n=0,\pm
1,\pm 2,\ldots $, are avoided by semicircles $C_{\rho }(\gamma _{n})$ in the
right half-plane with centers at $\gamma =\gamma _{n}$ and with small radius
$\rho $. We get the following terms: the sum of the integrals over the
straight segments of the positive and negative imaginary semiaxes between
the poles $\gamma _{n}$ and the sum of the integrals over the semicircles $%
C_{\rho }(\gamma _{n})$. In the limit $\rho \rightarrow 0$ the sum of the
integrals over the straight segments gives the principal values of the
integrals over the positive and negative imaginary semiaxes (denoted here as
p.v.). The integrals over the intervals $(0,im)$ and $(0,-im)$ cancel each
other, whereas the integral over $(-im,-i\infty )$ is the complex conjugate
of the integral over $(im,i\infty )$. For the sum of the integrals along the
semicircles $C_{\rho }(\gamma _{n})$, $n=0,1,2,\ldots $, one gets%
\begin{eqnarray}
&&\sum_{n=0}^{\infty }\int_{C(\gamma _{n})}d\gamma \frac{\bar{J}_{\beta
_{j}}^{(\lambda )}(\gamma a)}{\bar{H}_{\beta _{j}}^{(1,\lambda )}(\gamma a)}%
\left[ (E+\lambda sm)H_{\beta _{j}}^{(1)2}(\gamma r)-(E-\lambda sm)H_{\beta
_{j}+\epsilon _{j}}^{(1)2}(\gamma r)\right] \frac{\gamma /E}{e^{\beta E}+1}%
=-2T  \notag \\
&&\times \sum_{n=0}^{\infty }\frac{\bar{I}_{\beta _{j}}^{(\lambda )}(u_{0n}a)%
}{\bar{K}_{\beta _{j}}^{(\lambda )}(u_{0n}a)}\left[ (i\pi \left( 2n+1\right)
T+\lambda sm)K_{\beta _{j}}^{2}(u_{0n}r)+(i\pi \left( 2n+1\right) T-\lambda
sm)K_{\beta _{j}+\epsilon _{j}}^{2}(u_{0n}r)\right] ,  \label{Cn}
\end{eqnarray}%
where the notations for the modified Bessel functions with the bar are
defined in accordance of (\ref{FbIK}) with $\sqrt{\left( e^{\pi
i/2}u_{0n}a\right) ^{2}+m_{a}^{2}}=i\pi \left( 2n+1\right) Ta$. It can be
seen that the sum of the integrals for $C_{\rho }(\gamma _{n})$ with $%
n=\ldots ,-2,-1$ is the complex conjugate of the right-hand side of (\ref{Cn}%
). Introducing the modified Bessel functions in the integral over $%
(im,i\infty )$ we get%
\begin{eqnarray}
\langle \bar{\psi}\psi \rangle _{T\lambda }^{(b)} &=&\lambda \frac{1}{\pi
\phi _{0}}\sum_{j}\mathrm{p.v.}\int_{m}^{\infty }dx\,\mathrm{Im}\left\{
\frac{x}{e^{i\beta \sqrt{x^{2}-m^{2}}}+1}\frac{\bar{I}_{\beta
_{j}}^{(\lambda )}(xa)}{\bar{K}_{\beta _{j}}^{(\lambda )}(xa)}\right.  \notag
\\
&&\times \left[ \left( 1-\frac{\lambda ism}{\sqrt{x^{2}-m^{2}}}\right)
K_{\beta _{j}}^{2}(xr)+\left( 1+\frac{\lambda ism}{\sqrt{x^{2}-m^{2}}}%
\right) K_{\beta _{j}+\epsilon _{j}}^{2}(xr)\right]  \notag \\
&&-\lambda \frac{T}{\phi _{0}}\sum_{j}\sum_{n=0}^{\infty }\mathrm{Im}\left\{
\frac{\bar{I}_{\beta _{j}}^{(\lambda )}(u_{0n}a)}{\bar{K}_{\beta
_{j}}^{(\lambda )}(u_{0n}a)}\left[ (\pi \left( 2n+1\right) T-\lambda
ism)K_{\beta _{j}}^{2}(u_{0n}r)\right. \right.  \notag \\
&&\left. \left. +(\pi \left( 2n+1\right) T+\lambda ism)K_{\beta
_{j}+\epsilon _{j}}^{2}(u_{0n}r)\right] \right\} .  \label{FCb20}
\end{eqnarray}%
Now, by using the relations $\bar{I}_{\beta _{j}}^{(-)}(z)=\bar{I}_{\beta
_{j}}^{(+)\ast }(z)$, $\bar{K}_{\beta _{j}}^{(-)}(z)=\bar{K}_{\beta
_{j}}^{(+)\ast }(z)$, we can see that the contributions from the first term
in the right-hand side of (\ref{FCb20}) to the sum $\langle \bar{\psi}\psi
\rangle _{T}^{(b)}=\sum_{\lambda =\pm }\langle \bar{\psi}\psi \rangle
_{T\lambda }^{(b)}$ give $-\langle \bar{\psi}\psi \rangle _{\mathrm{vac}%
}^{(b)}$ and one finds%
\begin{eqnarray}
\langle \bar{\psi}\psi \rangle _{T}^{(b)} &=&-\frac{2T}{\phi _{0}}%
\sum_{j}\sum_{n=0}^{\infty }\mathrm{Im}\left\{ \frac{\bar{I}_{\beta
_{j}}(u_{0n}a)}{\bar{K}_{\beta _{j}}(u_{0n}a)}\left[ (\pi \left( 2n+1\right)
T-ism)K_{\beta _{j}}^{2}(u_{0n}r)\right. \right.  \notag \\
&&\left. \left. +(\pi \left( 2n+1\right) T+ism)K_{\beta _{j}+\epsilon
_{j}}^{2}(u_{0n}r)\right] \right\} -\langle \bar{\psi}\psi \rangle _{\mathrm{%
vac}}^{(b)}.  \label{FCb40}
\end{eqnarray}%
For the total boundary-induced FC (\ref{FCdec2}) the last term on the right
of (\ref{FCb40}) is cancelled by the boundary-induced part in the vacuum FC,
$\langle \bar{\psi}\psi \rangle _{\mathrm{vac}}^{(b)}$, and we get the
representation (\ref{FCbtotmu0}) for the boundary-induced FC in the case of
zero chemical potential. The corresponding formula is also obtained from (%
\ref{FCb4}) in the limit $\mu \rightarrow 0$.

Now let us consider the interior region, $r<a$. The thermal contributions to
the FC coming from particles and antiparticles are given by (\ref{FCTi})
with $\mu =0$. For the zero chemical potential the procedure we have used to
obtain the summation formula (\ref{SumAP}) from the generalized Abel-Plana
formula should be modified by taking into account that the function in the
integrand has poles $z=\pm i\gamma _{n}=\pm iu_{0n}a$, $n=0,1,2,\ldots $, on
the imaginary axis, corresponding to the zeros of $e^{\beta \sqrt{%
z^{2}/a^{2}+m^{2}}}+1$. In the part of the generalized Abel-Plana formula
corresponding to the integral along the imaginary axis these poles are
avoided by the semicircles $C_{\rho }(\gamma _{n})$. In the limit $\rho
\rightarrow 0$ we get%
\begin{eqnarray}
\sum_{l=1}^{\infty }T_{\beta _{j}}(\gamma _{j,l}^{(\lambda )})f(\gamma
_{j,l}^{(\lambda )}) &=&\int_{0}^{\infty }dx\,f(x)+\frac{\pi }{2}\underset{%
z=0}{\mathrm{Res}}\frac{\tilde{Y}_{\beta _{j}}^{(\lambda )}(z)}{\tilde{J}%
_{\beta _{j}}^{(\lambda )}(z)}f(z)  \notag \\
&&-2\sum_{n=0}^{\infty }\mathrm{Re}\left[ e^{-i\pi \beta _{j}}\frac{\tilde{K}%
_{\beta _{j}}^{(\lambda )}(u_{0n})}{\tilde{I}_{\beta _{j}}^{(\lambda
)}(u_{0n})}\underset{z=iu_{0n}}{\mathrm{Res}}f(z)\right]  \notag \\
&&-\frac{2}{\pi }\mathrm{p.v.}\int_{0}^{\infty }dx\,\mathrm{Re}\left[
e^{-i\pi \beta _{j}}f(xe^{\pi i/2})\frac{\tilde{K}_{\beta _{j}}^{(\lambda
)}(x)}{\tilde{I}_{\beta _{j}}^{(\lambda )}(x)}\right] .  \label{SumAP0}
\end{eqnarray}%
Note that in applying this formula to the FC the term coming from the poles $%
iu_{0n}$ is present for both $\lambda =+$ and $\lambda =-$, whereas in (\ref%
{SumAP}) the pole term is present only in case $\lambda \mu >0$. Further
transformations of the FC are similar to those for the exterior region. In
the region $r<a$, the expressions for $\langle \bar{\psi}\psi \rangle
_{T\lambda }^{(b)}$ and $\langle \bar{\psi}\psi \rangle _{T}^{(b)}$ are
obtained from (\ref{FCb20}) and (\ref{FCb40}) by the replacements $%
I\rightleftarrows K$, $\bar{I}\rightarrow \tilde{I}$ and $\bar{K}\rightarrow
\tilde{K}$. We see that the expression for $\langle \bar{\psi}\psi \rangle
_{b}$ is also directly obtained from (\ref{FCbTi}) in the limit $\mu
\rightarrow 0$.

\section{Zero temperature limit}

\label{sec:AppB}

In this section we consider the zero temperature limit of the expressions
for the FC obtained above. First of all for $|\mu |<m$ from (\ref{FCpm}) it
follows that $\lim_{T\rightarrow 0}\langle \bar{\psi}\psi \rangle =\langle
\bar{\psi}\psi \rangle _{\mathrm{vac}}$ and the FC coincides with that for
the vacuum state. In the case $|\mu |>m$ and for the exterior region one
gets
\begin{eqnarray}
\lim_{T\rightarrow 0}\langle \bar{\psi}\psi \rangle &=&\langle \bar{\psi}%
\psi \rangle _{\mathrm{vac}}+\lambda \frac{1}{2\phi _{0}}\sum_{j}\int_{0}^{%
\sqrt{\mu ^{2}-m^{2}}}d\gamma \frac{\gamma }{E}  \notag \\
&&\times \frac{(E+\lambda sm)g_{\beta _{j},\beta _{j}}^{(\lambda )2}(\gamma
a,\gamma r)-(E-\lambda sm)g_{\beta _{j},\beta _{j}+\epsilon _{j}}^{(\lambda
)2}(\gamma a,\gamma r)}{\bar{J}_{\beta _{j}}^{(\lambda )2}(\gamma a)+\bar{Y}%
_{\beta _{j}}^{(\lambda )2}(\gamma a)},  \label{FCT0}
\end{eqnarray}%
where $\lambda =+$ for $\mu >0$ and $\lambda =-$ for $\mu <0$ ($\mu =\lambda
|\mu |$). For the interior region%
\begin{equation}
\langle \bar{\psi}\psi \rangle =\langle \bar{\psi}\psi \rangle _{\mathrm{vac}%
}+\lambda \frac{a^{-2}}{2\phi _{0}}\sum_{j}\sum_{l=1}^{l_{m}}T_{\beta
_{j}}(\gamma _{j,l}^{(\lambda )})g(\gamma _{j,l}^{(\lambda )}),
\label{FCT0i}
\end{equation}%
with the same $\lambda $ as in (\ref{FCT0}) and $l_{m}$ defined by%
\begin{equation}
\gamma _{j,l_{m}}^{(\lambda )}\leq \sqrt{\mu ^{2}-m^{2}}<\gamma
_{j,l_{m}+1}^{(\lambda )}.  \label{poles}
\end{equation}%
The last terms in (\ref{FCT0}) and (\ref{FCT0i}) come from particles for $%
\mu >0$ and antiparticles for $\mu <0$. They occupy the states with energies
$E\leq |\mu |$.

Now let us consider the limit $T\rightarrow 0$ for the boundary-induced
contribution of the FC in the exterior region on the base of the formula (%
\ref{FCb4}). For small temperatures the dominant contribution to the sum
over $n$ in (\ref{FCb4}) comes from large values $n$ and, to the leading
order, we can replace the summation by the integration. The leading term
does not depend on temperature and is presented as%
\begin{equation}
\langle \bar{\psi}\psi \rangle ^{(b)}\approx -\frac{1}{\pi \phi _{0}}\sum_{j}%
\mathrm{Im}\left\{ \int_{-i\mu }^{\infty -i\mu }dx\,\frac{\bar{I}_{\beta
_{j}}(ua)}{\bar{K}_{\beta _{j}}(ua)}\left[ (x-ism)K_{\beta
_{j}}^{2}(ur)+(x+ism)K_{\beta _{j}+\epsilon _{j}}^{2}(ur)\right] \right\} ,
\label{GCbT01}
\end{equation}%
where $u=(x^{2}+m^{2})^{1/2}$. The integral in the right-hand side can be
written as the sum of two integrals: $\int_{-i\mu }^{\infty -i\mu
}dx=\int_{0}^{\infty }dx+\int_{-i\mu }^{0}dx$. The part in the FC with the
integral $\int_{0}^{\infty }dx$ coincides with the boundary-induced FC in
the vacuum state, . For $|\mu |<m$, introducing in the integral $\int_{-i\mu
}^{0}dx$ the integration variable $y=-ix$, we can see that the corresponding
integral under the imaginary sign is real and, hence, the contribution of
the integral $\int_{-i\mu }^{0}dx$ to the FC is zero. Consequently, we get $%
\lim_{T\rightarrow 0}\langle \bar{\psi}\psi \rangle ^{(b)}=\langle \bar{\psi}%
\psi \rangle _{\mathrm{vac}}^{(b)}$.

For $|\mu |>m$, we decompose the second integral as $\int_{-i\mu
}^{0}dx=\int_{-i\mu }^{-\lambda im}dx+\int_{-\lambda im}^{0}dx$, where $%
\lambda $ is defined by the relation $\mu =\lambda |\mu |$. The contribution
of part with the integral $\int_{-\lambda im}^{0}dx$ to the FC is zero by
the same reason as that for the integral $\int_{-i\mu }^{0}dx$ in the case $%
|\mu |<m$. In the integral $\int_{-i\mu }^{-\lambda im}dx$ we introduce $y$
in accordance with $x=-\lambda iy$ and then pass to a new integration
variable $z=\sqrt{y^{2}-m^{2}}$. Introducing the Bessel and Hankel functions
instead of the modified Bessel functions, the limiting value of the
boundary-induced FC is presented in the form%
\begin{eqnarray}
\lim_{T\rightarrow 0}\langle \bar{\psi}\psi \rangle ^{(b)} &=&\langle \bar{%
\psi}\psi \rangle _{\mathrm{vac}}^{(b)}-\lambda \frac{1}{2\phi _{0}}\sum_{j}%
\mathrm{Re}\left\{ \int_{0}^{\sqrt{\mu ^{2}-m^{2}}}dz\,z\,\frac{\bar{J}%
_{\beta _{j}}^{(\lambda )}(za)}{\bar{H}_{\beta _{j}}^{(l,\lambda )}(za)}%
\right.  \notag \\
&&\times \left. \left[ \left( 1+\frac{\lambda sm}{\sqrt{z^{2}+m^{2}}}\right)
H_{\beta _{j}}^{(l)2}(zr)-\left( 1-\frac{\lambda sm}{\sqrt{z^{2}+m^{2}}}%
\right) H_{\beta _{j}+\epsilon _{j}}^{(l)2}(zr)\right] \right\} ,
\label{FCbT0}
\end{eqnarray}%
where $\mu =\lambda |\mu |$, $l=1$ for $\lambda =+$, $l=2$ for $\lambda =-$,
and we use the notation (\ref{Fbar}). This coincides with the
boundary-induced part obtained from (\ref{FCT0}) by using the relation (\ref%
{ident1}).

Now we consider the zero temperature limit in the interior region, based on
the representation (\ref{FCbTi}) for the boundary-induced part. To the
leading order, we replace the summation over $n$ by integration with the
result%
\begin{equation}
\lim_{T\rightarrow 0}\langle \bar{\psi}\psi \rangle ^{(b)}=-\frac{1}{\pi
\phi _{0}}\sum_{j}\mathrm{Im}\left\{ \int_{-i\mu }^{\infty -i\mu }dx\,\frac{%
\tilde{K}_{\beta _{j}}(ua)}{\tilde{I}_{\beta _{j}}(ua)}\left[
(x-ism)I_{\beta _{j}}^{2}(ur)+(x+ism)I_{\beta _{j}+\epsilon _{j}}^{2}(ur)%
\right] \right\} ,  \label{FCbT0i}
\end{equation}%
where $u=(x^{2}+m^{2})^{1/2}$. Similar to the case of the exterior region,
we split the integral as $\int_{-i\mu }^{\infty -i\mu }dx=\int_{0}^{\infty
}dx+\int_{-i\mu }^{0}dx$. The part in the FC corresponding to the integral
over $[0,\infty )$ gives $\langle \bar{\psi}\psi \rangle _{\mathrm{vac}%
}^{(b)}$ (see (\ref{FCbvaci})). For $|\mu |<m$ and for the integral over $%
[-i\mu ,0]$, in the arguments of the modified Bessel functions $u$ is
positive. Introducing a new integration variable $y=i\lambda x$, we see that
the integral is real and does no contribute to (\ref{FCbT0i}). Hence, for $%
|\mu |<m$, again we get $\lim_{T\rightarrow 0}\langle \bar{\psi}\psi \rangle
^{(b)}=\langle \bar{\psi}\psi \rangle _{\mathrm{vac}}^{(b)}$. In the case $%
|\mu |>m$, the nonzero contribution comes from the part of the integral over
$[-i\mu ,-\lambda im]$. In addition we should take into account that the
integrand in (\ref{FCbT0i}) may have poles corresponding to $ua=-\lambda
i\gamma _{j,l}^{(\lambda )}$, $l=1,2,\ldots ,l_{m}$, (defined by (\ref{poles}%
)) in that segment of the imaginary axis. Passing to a new integration
variable $z=\lambda i(x^{2}+m^{2})^{1/2}$, we avoid possible poles $z=\gamma
_{j,l}^{(\lambda )}/a$ by small semicircles $C_{\rho }(\gamma
_{j,l}^{(\lambda )}/a)$ in the right-half plane with a small radius $\rho $
and with the center at $z=\gamma _{j,l}^{(\lambda )}/a$. In the limit $\rho
\rightarrow 0$, the sum of the integrals over the straight segments between
the poles gives the principal value of the integral and we get the following
representation
\begin{eqnarray}
\lim_{T\rightarrow 0}\langle \bar{\psi}\psi \rangle ^{(b)} &=&\langle \bar{%
\psi}\psi \rangle _{\mathrm{vac}}^{(b)}-\frac{1}{\pi \phi _{0}}\sum_{j}%
\mathrm{Im}\left\{ \left[ \mathrm{p.v.}\int_{0}^{\sqrt{\mu ^{2}-m^{2}}%
}dz-\sum_{l=1}^{l_{m}}\int_{C_{\rho }(\gamma _{j,l}^{(\lambda )}/a)}dz\right]
z\,\frac{\tilde{K}_{\beta _{j}}(ua)}{\tilde{I}_{\beta _{j}}(ua)}\right.
\notag \\
&&\left. \times \left[ (1+\frac{\lambda sm}{\sqrt{z^{2}+m^{2}}})I_{\beta
_{j}}^{2}(ur)+(1-\frac{\lambda sm}{\sqrt{z^{2}+m^{2}}})I_{\beta
_{j}+\epsilon _{j}}^{2}(ur)\right] _{u=-\lambda iz}\right\} .
\label{FCbT0i2}
\end{eqnarray}%
By using the relations $\tilde{J}_{\beta _{j}}^{(\lambda )\prime
}(w)=-2/[T_{\beta _{j}}^{(\lambda )}(w)J_{\beta _{j}}(w)]$ and $\widetilde{Y}%
_{\beta _{j}}^{(\lambda )}(x)=2/[\pi J_{\beta _{j}}(x)]$, valid for $%
w=\gamma _{j,l}^{(\lambda )}$, it can be shown that the integral over $%
C_{\rho }(\gamma _{j,l}^{(\lambda )}/a)$ is equal to $\lambda i\pi T_{\beta
_{j}}^{(\lambda )}(w)g(w)/(2a^{2})$. Introducing in the integral over $[0,%
\sqrt{\mu ^{2}-m^{2}}]$ the Bessel and Hankel functions, the integrand
becomes $\widetilde{H}_{\beta _{j}}^{(l,\lambda )}(za)g(za)/\tilde{J}_{\beta
_{j}}^{(\lambda )}(za)$ with the real part $g(za)$. As a result, the
zero-temperature limit is presented as
\begin{eqnarray}
\lim_{T\rightarrow 0}\langle \bar{\psi}\psi \rangle ^{(b)} &=&\langle \bar{%
\psi}\psi \rangle _{\mathrm{vac}}^{(b)}-\lambda \frac{1}{2\phi _{0}a}%
\sum_{j}\int_{0}^{\sqrt{\mu ^{2}-m^{2}}}dz\,g(za)  \notag \\
&&+\lambda \frac{a^{-2}}{2\phi _{0}}\sum_{j}\sum_{l=1}^{l_{m}}\,T_{\beta
_{j}}^{(\lambda )}(\gamma _{j,l}^{(\lambda )})g(\gamma _{j,l}^{(\lambda )}).
\label{FCbT0i3}
\end{eqnarray}%
From (\ref{gzi}) it follows that $g(za)/a$ does not depend on $a$. By taking
into account that for the boundary-free geometry one has%
\begin{equation}
\lim_{T\rightarrow 0}\langle \bar{\psi}\psi \rangle ^{(0)}=\langle \bar{\psi}%
\psi \rangle _{\mathrm{vac}}^{(0)}+\lambda \frac{1}{2\phi _{0}a}%
\sum_{j}\int_{0}^{\sqrt{\mu ^{2}-m^{2}}}dz\,g(za),  \label{FCT0i2}
\end{equation}%
we see that in the zero-temperature limit for the total FC the integral term
in (\ref{FCbT0i3}) is cancelled by the last term in (\ref{FCT0i2}) and the
formula (\ref{FCT0i}) is obtained. Hence, we have shown that both the
representations for exterior and interior FCs give the same zero-temperature
limit.

\end{document}